\documentclass[aps, prb, twocolumn, superscriptaddress, amsmath,  tightenlines, longbibliography]{revtex4-1}

\usepackage{dcolumn}
\usepackage{graphicx}
\usepackage{mathrsfs}
\usepackage{subfigure}
\usepackage{booktabs}
\usepackage{amsmath}
\usepackage{physics}
\usepackage{dsfont}
\usepackage{amstext}
\usepackage{amssymb}
\usepackage{amsbsy}
\usepackage{bbm}
\usepackage{amsthm}
\usepackage{graphicx}
\usepackage{color}
\usepackage[colorlinks,citecolor=blue]{hyperref}

\usepackage{url}
\usepackage[colorlinks]{hyperref}
\hypersetup{%
	plainpages=true,
	breaklinks=true,       
	hypertexnames=false,  
	pageanchor=true,
	colorlinks=true,
	linkcolor={blue},
	citecolor={red},
	urlcolor={blue},
	anchorcolor={black}
}

 \makeatletter

\newcommand{\Rmnum}[1]{\expandafter\@slowromancap\romannumeral #1@}
\makeatother

\begin{document}

\title{Composite-State Localization Beyond the External Landscape in \\ Non-Hermitian Quasicrystals}
\author{Tian Zhou}
\thanks{These authors contributed equally}
\affiliation{School of Physics and Optoelectronics, South China University of Technology,  Guangzhou 510640, China}
\author{Xue-Bin Wang}
\thanks{These authors contributed equally}
\affiliation{School of Physics and Optoelectronics, South China University of Technology,  Guangzhou 510640, China}
\author{Zhongmin Yang}
\email[E-mail: ]{yangzm@scut.edu.cn}
\affiliation{School of Physics and Optoelectronics, South China University of Technology, Guangzhou 510640, China}
\affiliation{State Key Laboratory of Luminescent Materials and Devices and Institute of Optical Communication Materials, South China University of Technology, Guangzhou 510640, China}
\author{Tao Liu}
\email[E-mail: ]{liutao0716@scut.edu.cn}
\affiliation{School of Physics and Optoelectronics, South China University of Technology,  Guangzhou 510640, China}

\date{{\small \today}}


\begin{abstract}
A composite excitation need not inherit the localization behavior of its
constituents. We show that an interacting non-Hermitian quasiperiodic
ladder realizes a controllable and reversible localization inversion
between composite and unbound excitations, where internal configuration,
rather than only the external potential, becomes a control parameter for
localization. Opposite complex potentials on the two legs cancel at first
order for a same-rung pair but act directly on separated particles,
allowing extended composite states to persist while the unpaired sector
becomes localized. A strong-coupling theory identifies the composite state
as an emergent weakly modulated non-Hermitian quasicrystal generated by
virtual unpaired configurations. Breaking the potential antisymmetry
restores a direct modulation of the composite band and reverses the
localization hierarchy. Engineering configuration-space pathways further
stabilizes an extended composite band embedded within a localized
continuum, the inverse of the conventional bound-state-in-the-continuum
scenario. Our results establish internal configuration as a reversible
control parameter for localization.
\end{abstract}

\maketitle

\section{Introduction}

 The conventional paradigm of Anderson localization attributes transport
 suppression to interference induced by a random or quasiperiodic spatial
 landscape
 \cite{PhysRev.109.1492,Pierce1993,PhysRevLett.103.013901,Thouless1974}.
 Interactions, however, can fundamentally alter this single-particle
 picture. When particles form a composite excitation, such as a doublon
 \cite{Winkler2006}, the resulting object is not simply a heavier particle
 with a renormalized hopping amplitude
 \cite{PhysRevLett.73.2607}. Its internal structure modifies the virtual
 processes governing its motion and can generate an effective kinetic and
 potential landscape distinct from that experienced by its constituents.
 
 Quasiperiodic lattices provide a highly controllable platform for
 exploring such configuration-dependent localization. Their deterministic
 long-range order supports extended, critical, localized, and mobility-edge
 regimes without genuine randomness
 \cite{PhysRevLett.53.2477,Goblot2020,PhysRevLett.125.196604,
 	PhysRevB.106.L140203,Wang2024}. These phenomena are further enriched in
 non-Hermitian systems \cite{ShunyuYao2018,PhysRevLett.123.066404,PhysRevLett.125.126402, PhysRevLett.122.076801, PhysRevLett.121.026808,Leefmans2022,arXiv:1802.07964,  PhysRevLett.123.016805, PhysRevLett.123.206404,PhysRevLett.123.206404, PhysRevX.9.041015, PhysRevLett.124.086801, PhysRevLett.127.196801, zf4k-ytgt, PhysRevLett.129.093001,  vxgf-59xt, PhysRevLett.134.176601,lpm2-vcb4}, where complex potentials generate unconventional
 localization transitions, complex spectral evolution, and mixed
 localization regimes
 \cite{PhysRevB.100.054301,PhysRevB.101.174205,PhysRevLett.122.237601,
 	PhysRevA.103.033325,PhysRevB.102.024205,Weidemann2022,Lin2022,tz2n-lqxx}.
 Interactions introduce an additional dependence on particle configuration:
 two particles occupying the same unit cell need not experience the same
 effective landscape as two spatially separated particles. A fundamental
 question is therefore whether a composite excitation can be engineered to
 exhibit a localization response opposite to that of its constituents
 within the same microscopic environment.
 
 Here, we answer this question by demonstrating a controllable
 configuration-selective localization inversion in an interacting
 non-Hermitian quasiperiodic ladder. Opposite complex potentials on the two
 legs act directly on spatially separated particles but cancel at first
 order for same-rung composite excitations. Their motion is instead
 generated through virtual unpaired configurations, producing an emergent
 weakly modulated non-Hermitian quasicrystal. This mechanism allows extended
 composite states to persist while the unpaired sector becomes localized.
 Breaking the potential antisymmetry restores a direct quasiperiodic
 modulation of the composite band and reverses the localization hierarchy.
 Furthermore, by engineering the configuration-space pathways responsible
 for composite-state propagation, we stabilize an extended composite band
 embedded within a localized continuum of unpaired states, providing an
 inverse counterpart to the conventional bound state in the continuum
\cite{PhysRevLett.133.193001,PhysRevLett.133.140202}. Our results
 establish internal configuration as an independent and reversible control
 parameter for localization, opening a route toward selective confinement,
 transport, and spectroscopy of composite and elementary excitations in
 synthetic classical and quantum platforms.
 
The remainder of the paper is organized as follows. In Sec.~II, we introduce the interacting non-Hermitian quasiperiodic ladder and the real-space localization diagnostics used throughout the work. In Sec.~III, we demonstrate the localization inversion under antisymmetric leg potentials and derive the effective Hamiltonian governing the composite-pair sector. In Sec.~IV, we show that breaking the potential antisymmetry restores a direct quasiperiodic modulation of the composite band and reverses the localization hierarchy. In Sec.~V, we introduce an interaction-engineered configuration-space pathway that stabilizes an extended composite band embedded within a localized continuum. Finally, in Sec.~VI, we summarize our main results and discuss their broader implications.

\section{Model}

\begin{figure}[!tb]
	\centering
	\includegraphics[width=8.4cm]{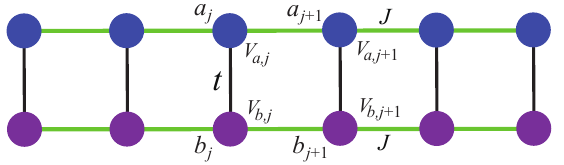}
	\caption{Interacting non-Hermitian quasiperiodic ladder. Hard-core 	bosons hop along the two legs with amplitude $J$ and between the legs		with amplitude $t$. Particles occupying the same rung interact with		strength $U_1$. The upper and lower legs experience the complex
		quasiperiodic potentials  	$V_{a,j} = V\cos(2\pi\alpha j+ih+\Delta\phi/2)$ and
		$V_{b,j} = -V\cos(2\pi\alpha j+ih-\Delta\phi/2)$, respectively. }\label{lattice_model}
\end{figure}

To realize localization inversion between composite pairs and unbound
particles, we consider an interacting non-Hermitian quasiperiodic ladder formed by two coupled chains, as illustrated in Fig.~\ref{lattice_model}. Each leg hosts hard-core bosons, while particles occupying the same rung interact. The two legs experience nearly opposite complex quasiperiodic
potentials, which act differently on spatially separated particles and
same-rung pairs and thereby generate configuration-dependent
localization landscapes. The Hamiltonian is $\mathcal{H}_1 =\mathcal{H}_0+\mathcal{V}+\mathcal{H}_{\mathrm{int}}$,
with
\begin{align}\label{H1}
	\mathcal{H}_0  =   \sum_{j} \left(J  a_{j+1}^\dagger  a_{j} + J  b_{j+1}^\dagger  b_{j}+ t  a_{j}^\dagger  b_{j} + \textrm{H.c.} \right),  
\end{align}
\begin{align}\label{Hpot}
	\mathcal{V}(\Delta\phi)
	=~
	V\sum_j
	\Big[
	&\cos(2\pi\alpha j+ih+\Delta\phi/2)n_{a,j}
	\nonumber\\
	&-\cos(2\pi\alpha j+ih-\Delta\phi/2)n_{b,j}
	\Big],
\end{align}
\begin{align}\label{Hint}
	\mathcal{H}_\text{int}  =   U_1 \sum_{j} n_{a,j} n_{b,j}.
\end{align}
Here, $a_j^{\dagger}$ and $b_j^{\dagger}$ create hard-core bosons on rung $j$ of the upper and lower legs, respectively, and $n_{\beta,j}=\beta_j^{\dagger}\beta_j$ for $\beta=a,b$. The parameters
$J$ and $t$ denote the intraleg and interleg hopping amplitudes, and $U_1$
is the interaction strength of a same-rung pair. We choose the incommensurate wave number $\alpha=(\sqrt{5}-1)/2$. Non-Hermiticity is
introduced through the imaginary phase shift $ih$, with $h$ controlling
the complex deformation of the onsite potential \cite{PhysRevLett.122.237601}. The relative phase $\Delta\phi$ controls the departure from exact antisymmetry between the two legs. For $\Delta\phi=0$, the onsite contributions cancel exactly at first
order for two particles occupying the same rung. A finite
$\Delta\phi$ makes this first-order cancellation incomplete, producing
the residual pair potential
$-2V\sin(2\pi\alpha j +ih)\sin(\Delta\phi/2)$.  Since $\mathcal{H}_1$ conserves the total
particle number, we focus on the two-excitation sector. Unless stated
otherwise, periodic boundary conditions are imposed.

To characterize the real-space localization of both paired and unpaired
eigenstates on an equal footing, we define the normalized rung density
of the $m$th right eigenstate $|\psi_m^R\rangle$ as 
\begin{equation}
	\rho_{j,m}= \frac{ \langle \psi_m^R|\left(n_{a,j}+n_{b,j}\right)|\psi_m^R\rangle}{2\langle \psi_m^R|\psi_m^R\rangle},
\end{equation}
 with $\sum_{j=1}^{L}\rho_{j,m}=1$. 
The corresponding real-space density inverse participation ratio is
defined as
\begin{equation}
	\mathrm{IPR}_{m}	=	\sum_{j=1}^{L}	\rho_{j,m}^{\,2}.	
\end{equation}\label{eq:density_ipr}
For both paired and unpaired states, $\mathrm{IPR}_{m}\propto L^{-1}$ ($L$ is length of ladder) indicates a spatially extended density distribution, whereas a finite value signals localization. For a strongly localized paired state, both particles occupy approximately the same rung and $\mathrm{IPR}_{m}\simeq 1$, whereas for two well-separated localized particles one typically has $\mathrm{IPR}_{m}\simeq 1/2$. Therefore, solely for visual comparison, the density IPR of the unpaired sector is multiplied by a factor of two in the figures below. This rescaling does not affect the identification of extended and localized states.

Before exploring the interacting regime, we establish the single-excitation
localization landscape of the non-Hermitian ladder. While an isolated
non-Hermitian quasicrystal exhibits a transition from extended to localized
states at $h_c=\ln(2J/V)$ \cite{PhysRevLett.122.237601}, the coupled ladder
with antisymmetric quasiperiodic potentials develops a richer sequence due
to interchain hybridization. As detailed in Appendix \ref{append1}, increasing the non-Hermitian strength drives
the single-particle spectrum from an extended phase ($h<h_1$), through a
mixed regime with coexisting extended and localized states
($h_1<h<h_2$), to a fully localized phase ($h>h_2$). Introducing a relative
phase difference $\Delta\phi$ between the two leg potentials breaks the
antisymmetric configuration and shifts the localization boundaries, while
preserving this overall transition sequence (see Appendix \ref{append1}). These single-particle results provide the reference
localization background for identifying the interaction-induced inversion
between composite pairs and unbound particles.

 \begin{figure*}[tb]
 	\centering
 	\includegraphics[width=18cm]{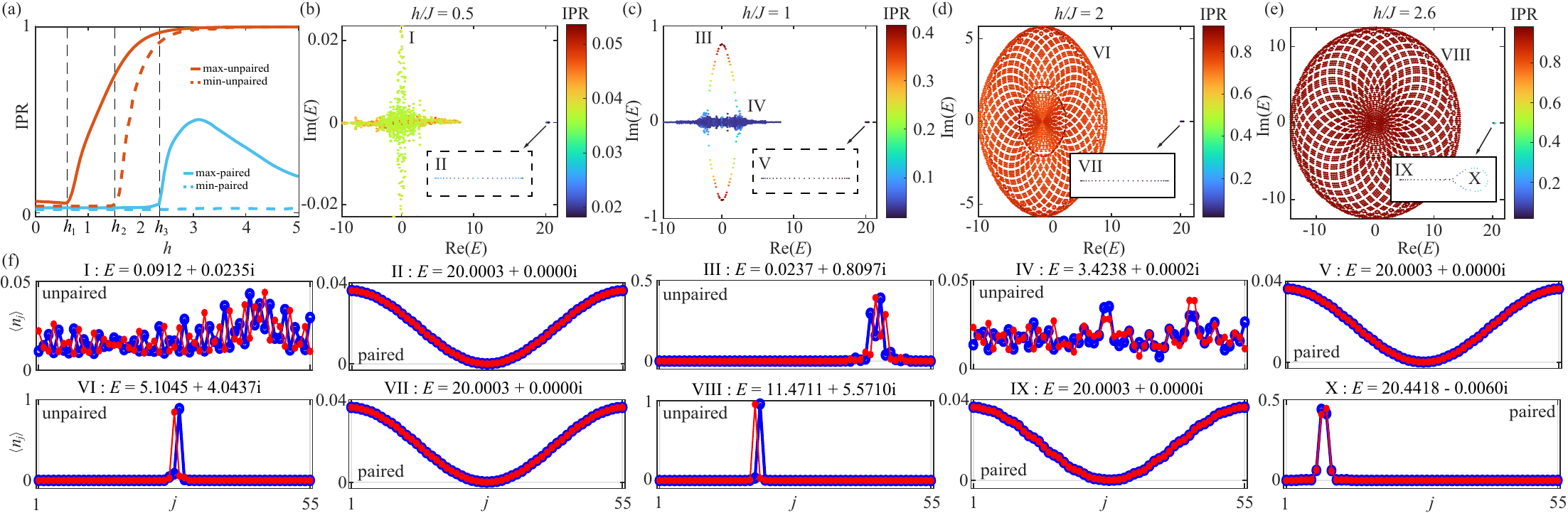}
 	\caption{(a) Evolution of the maximum (solid lines) and minimum (dashed lines) IPR of paired (light blue) and unpaired (vermilion) states as a function of the non-Hermitian strength $h$. Three characteristic localization transition points are identified: $h_1\simeq0.6$, $h_2\simeq1.58$, and $h_3\simeq2.37$. The first two transitions correspond to the successive extended--mixed--localized evolution of the unpaired states, while the third transition marks the onset of localization of the paired states. 		(b)-(e) Complex-energy spectra at representative values of $h$, with colors indicating the corresponding IPR values. The enlarged inset shows the energy region containing the paired states. 		(f) Representative particle-density distributions $\langle n_{\beta,j}\rangle$ for paired and unpaired states selected from the energy spectrum. The blue and red  curves denote the particle distributions on the top and bottom chains, respectively.  The parameters are $\Delta \phi = 0$, $U_1/J=20$, $V/J=1$, and $t/J=2$ with $L=55$. Unless otherwise specified, these parameters are used throughout the subsequent figures.
 	}\label{IPR_and_spectrum}
 \end{figure*}

\section{Localization Inversion between Composite Pairs and	Unbound Particles under Antisymmetric Potentials}

In the two-excitation sector, the interaction $U_1$ separates the Hilbert space into same-rung paired configurations and spatially separated
unpaired configurations. We identify paired states by their dominant
weight in the basis
$|\phi_j\rangle=a_j^\dagger b_j^\dagger|0\rangle$,
while unpaired states correspond to particles occupying different rungs.
This configuration dependence gives rise to distinct effective
localization landscapes, allowing composite pairs and their constituent
particles to exhibit opposite localization responses under the same
quasiperiodic potential. We first consider the antisymmetric case
$\Delta\phi=0$.

As shown in Fig.~\ref{IPR_and_spectrum}(a), the maximum and minimum IPR
values of paired and unpaired states evolve differently with increasing
non-Hermitian strength $h$, revealing distinct localization regimes.
The corresponding complex-energy spectra and representative density
profiles $\langle n_{\beta,j}\rangle=\bra{\psi_m^R}\beta_j^\dagger\beta_j\ket{\psi_m^R} $ ($\beta=a,b$) are shown in Fig.~\ref{IPR_and_spectrum}(b)--(f). The unpaired
states follow the localization behavior of the single-particle ladder:
they evolve from an extended phase ($h<h_1$), through a mixed regime with
coexisting extended and localized states ($h_1<h<h_2$), and finally into
a localized phase ($h>h_2$). This behavior reflects their direct
coupling to the complex quasiperiodic potential.

In contrast, as shown in Fig.~\ref{IPR_and_spectrum}(a, d)  and the corresponding particle-density distributions in (f), the paired states retain their extended character over a broad parameter regime ($h<h_3$). The origin of this localization contrast is revealed by the effective Hamiltonian of the paired sector. For two particles occupying the same rung, the first-order quasiperiodic contribution cancels because the two legs experience opposite onsite potentials, $V_{a,j}+V_{b,j}=0$, giving the bare pair energy $E_j^{(0)}=U_1+V_{a,j}+V_{b,j}=U_1$. Therefore, unlike unpaired particles, the composite pair does not
directly experience the original quasiperiodic landscape at first order.
 
The quasiperiodic modulation is not completely removed from the pair
dynamics, however. In the strongly interacting regime, pair motion is
generated by virtual excursions into the unpaired sector, resulting in
an emergent non-Hermitian quasicrystal described by
(see details in Appendix \ref{append2})
\begin{equation}
	H_{\rm pair}^{\rm eff}
	=
	\sum_j
	(\epsilon_j+U_1)
	|\phi_j\rangle\langle\phi_j|
	+
	\sum_j	
	J_j^{\rm eff}\left(	|\phi_{j+1}\rangle\langle\phi_j|
	+\mathrm{H.c.}
	\right).
\end{equation}
Here, the effective modulated hopping is
\begin{equation}
	J_j^{\rm eff}
	=
	\frac{J^2}
	{U_1+V_{a,j}-V_{a,j+1}}
	+
	\frac{J^2}
	{U_1+V_{b,j}-V_{b,j+1}},
\end{equation}
and the effective modulated onsite energy is
\begin{equation}
	\epsilon_j
	=	\sum_{\eta=\pm1}
	\left[
	\frac{J^2}
	{U_1+V_{a,j}-V_{a,j+\eta}}
	+
	\frac{J^2}
	{U_1+V_{b,j}-V_{b,j+\eta}}
	\right].
\end{equation}

Thus, the pair sector realizes a distinct effective non-Hermitian
quasicrystal generated by interaction-induced virtual processes. For
$U_1\gg |J|,|V|$, these corrections scale as $J^2/U_1$ and remain much
weaker than the direct quasiperiodic modulation experienced by unpaired
particles. Together with the hard-core constraint, which restricts the
available pair-breaking channels, and the interaction-induced spectral
separation of order $U_1$, this emergent weak landscape stabilizes
extended composite pairs even when the unpaired states have already
localized.

For sufficiently large non-Hermitian strength ($h>h_3$), the extended
pair regime gradually breaks down. The increasing complex quasiperiodic
modulation enhances the spatial variation of the effective parameters
$J_j^{\rm eff}$ and $\epsilon_j^{\rm eff}$, thereby strengthening the
effective non-Hermitian quasicrystal experienced by the composite pair.
Although the paired and unpaired sectors remain spectrally distinguishable,
the effective quasiperiodic modulation within the paired manifold becomes
strong enough to localize the composite excitation. Consequently, the
interaction-induced extended pair regime is eventually replaced by a
localized paired phase [see Fig.~\ref{IPR_and_spectrum}(a),(e) and the
corresponding density distributions in (f)]. Finite-size analysis confirms that the results are insensitive to the lattice size, while increasing $U_1$ broadens the parameter regime in which extended pair states persist alongside localized unpaired states (see details in Appendix \ref{append3}).

\section{Localization Inversion between Composite Pairs and Unbound Constituents}

We next demonstrate the opposite configuration-selective localization
regime, where the composite pair becomes localized while the unbound
particles remain extended. This is achieved by breaking the exact
antisymmetry between the quasiperiodic potentials on the two legs through
a relative phase offset $\Delta\phi$. For a rung-bound pair
$|\phi_j\rangle=a_j^\dagger b_j^\dagger|0\rangle$,
the first-order quasiperiodic contribution becomes
\begin{align}
	V_{j,\mathrm{pair}}^{(0)}
	=
	-2V\sin(2\pi\alpha j+ih)\sin(\Delta\phi/2).
\end{align}
At $\Delta\phi=0$, this term vanishes exactly and the pair energy reduces
to $E_{\mathrm{pair}}^{(0)}=U_1$. A finite phase offset restores a direct
quasiperiodic modulation in the paired sector, which for
$|\Delta\phi|\ll1$ appears as
\begin{align}
	V_{j,\mathrm{pair}}^{(0)}
	\simeq
	-V\Delta\phi\sin(2\pi\alpha j+ih).
\end{align}
Thus, even a weak breaking of the potential antisymmetry generates a
first-order quasiperiodic landscape for the composite excitation.

In addition to this direct contribution, the motion of the composite pair
is generated by virtual excursions into the unpaired sector. To second
order in the hopping $J$, these processes produce a spatially modulated
effective pair hopping (see details in Appendix \ref{append2}),
\begin{align}
	\mathcal{J}_j^{\rm eff}
	=
	\frac{J^2}{2}
	\Bigg[
	&
	\frac{1}{U_1+V_{a,j}-V_{a,j+1}}
	+
	\frac{1}{U_1+V_{a,j+1}-V_{a,j}}
	\nonumber\\
	&+
	\frac{1}{U_1+V_{b,j}-V_{b,j+1}}
	+
	\frac{1}{U_1+V_{b,j+1}-V_{b,j}}
	\Bigg],
\end{align}
together with a renormalized onsite energy
\begin{align}
	\epsilon_j^{\rm eff}
	= &	\sum_{\eta=\pm1}	\left[	\frac{J^2}{U_1+V_{a,j}-V_{a,j+\eta}}
	+	\frac{J^2}{U_1+V_{b,j}-V_{b,j+\eta}}	\right] \nonumber \\ & + 	E_j^{(0)},
\end{align}
where
\begin{align}
	E_{j}^{(0)}
	 = 	U_1	-	2V\sin(2\pi\alpha j+ih)	\sin(\Delta\phi/2).
\end{align}
The resulting effective pair Hamiltonian for $\Delta \phi \neq 0$ can therefore be written as
\begin{align}
	\mathcal{H}_{\rm pair}^{\rm eff}
	=\sum_j	\mathcal{J}_j^{\rm eff}	\left(	|\phi_{j+1}\rangle\langle\phi_j|
	++\mathrm{H.c.}	\right)+	\sum_j	\epsilon_j^{\rm eff}	|\phi_j\rangle\langle\phi_j|.
\end{align}
Thus, the composite sector experiences an emergent non-Hermitian
quasiperiodic landscape containing two distinct contributions: the direct
first-order modulation
$V_{a,j}+V_{b,j}
=-2V\sin(2\pi\alpha j+ih)\sin(\Delta\phi/2)$,
and the interaction-induced corrections
$\mathcal{J}_j^{\rm eff}$ and
$\epsilon_j^{\rm eff}-E_j^{(0)}\sim O(J^2/U_1)$
generated by virtual processes.

This effect is strongly amplified by the reduced kinetic scale of the
pair. Since pair propagation occurs through second-order virtual
processes, its effective bandwidth is
$\mathcal{J}_j^{\rm eff}\sim J^2/U_1$, much smaller than the single-particle
kinetic scales $J$ and $t$. Consequently, the restored first-order
quasiperiodic modulation can readily dominate the much narrower composite
band and localize the pair even when the unpaired particles remain
extended.

\begin{figure}[!tb]
	\centering
	\includegraphics[width=8.4cm]{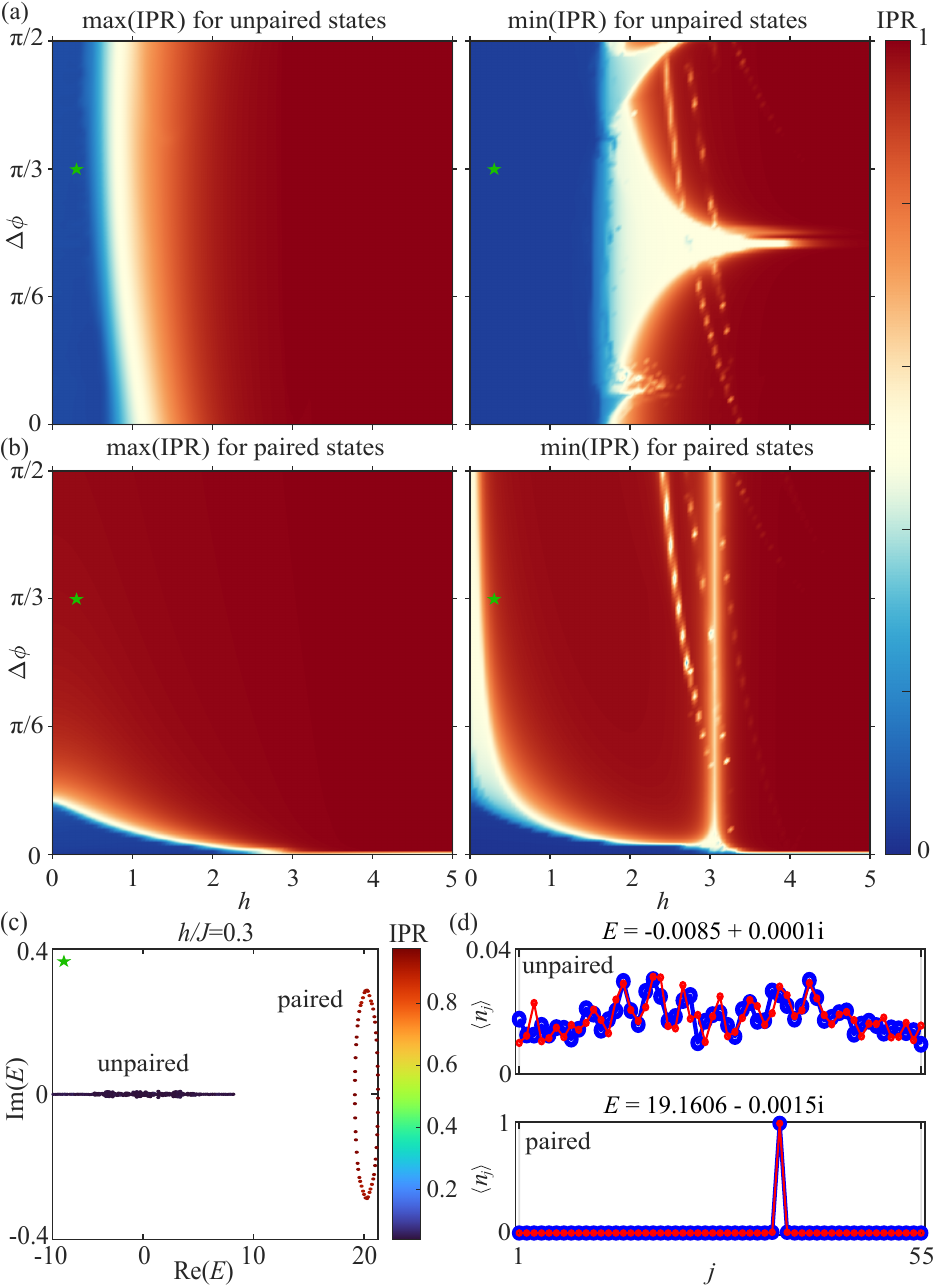}
	\caption{(a,b) Localization-delocalization phase diagrams in the $(h,\Delta\phi)$ parameter plane, characterized by the maximum and minimum IPRs of the unpaired  (a) and paired (b) states. The green star marks the representative parameters $h/J=0.3$ and $\Delta\phi=\pi/3$. (c) Complex-energy spectrum at the point marked in (a), with the color scale representing the IPR of each eigenstate.  (d) Representative particle-density distributions of the paired and unpaired eigenstates selected from (a,b). Blue  and red indicate the particle densities on top and bottom chains, respectively.
	}\label{IPR_and_spectrum_for_difV}
\end{figure}

The phase diagrams in Fig.~\ref{IPR_and_spectrum_for_difV}(a,b) reveal
this localization inversion. For $\Delta\phi=0$, the cancellation of the
first-order pair potential allows extended paired states to survive beyond
the localization transition of the unpaired sector. In contrast, finite
$\Delta\phi$ restores a direct quasiperiodic modulation in the narrow pair
band and reverses the localization hierarchy, driving the composite
excitation into a localized regime while the unpaired sector remains
delocalized. Thus, binding does not simply renormalize the localization
tendency; it generates a configuration-dependent effective landscape that
can reverse the localization response between a composite excitation and
the particles from which it is formed.

A representative example is shown for $h/J=0.3$ and
$\Delta\phi=\pi/3$. The complex-energy spectrum in
Fig.~\ref{IPR_and_spectrum_for_difV}(c) separates into a high-IPR paired
band and a low-IPR unpaired band, while the density distributions in
Fig.~\ref{IPR_and_spectrum_for_difV}(d) directly confirm the inverted localization hierarchy: localized
composite pairs accompanied by an extended unpaired sector. Combined
with the opposite configuration realized for antisymmetric potentials,
where extended pairs survive within localized unpaired states, these
results demonstrate a controllable reversal of the localization
hierarchy between a composite excitation and its constituents.

\begin{figure}[!tb]
	\centering
	\includegraphics[width=8.6cm]{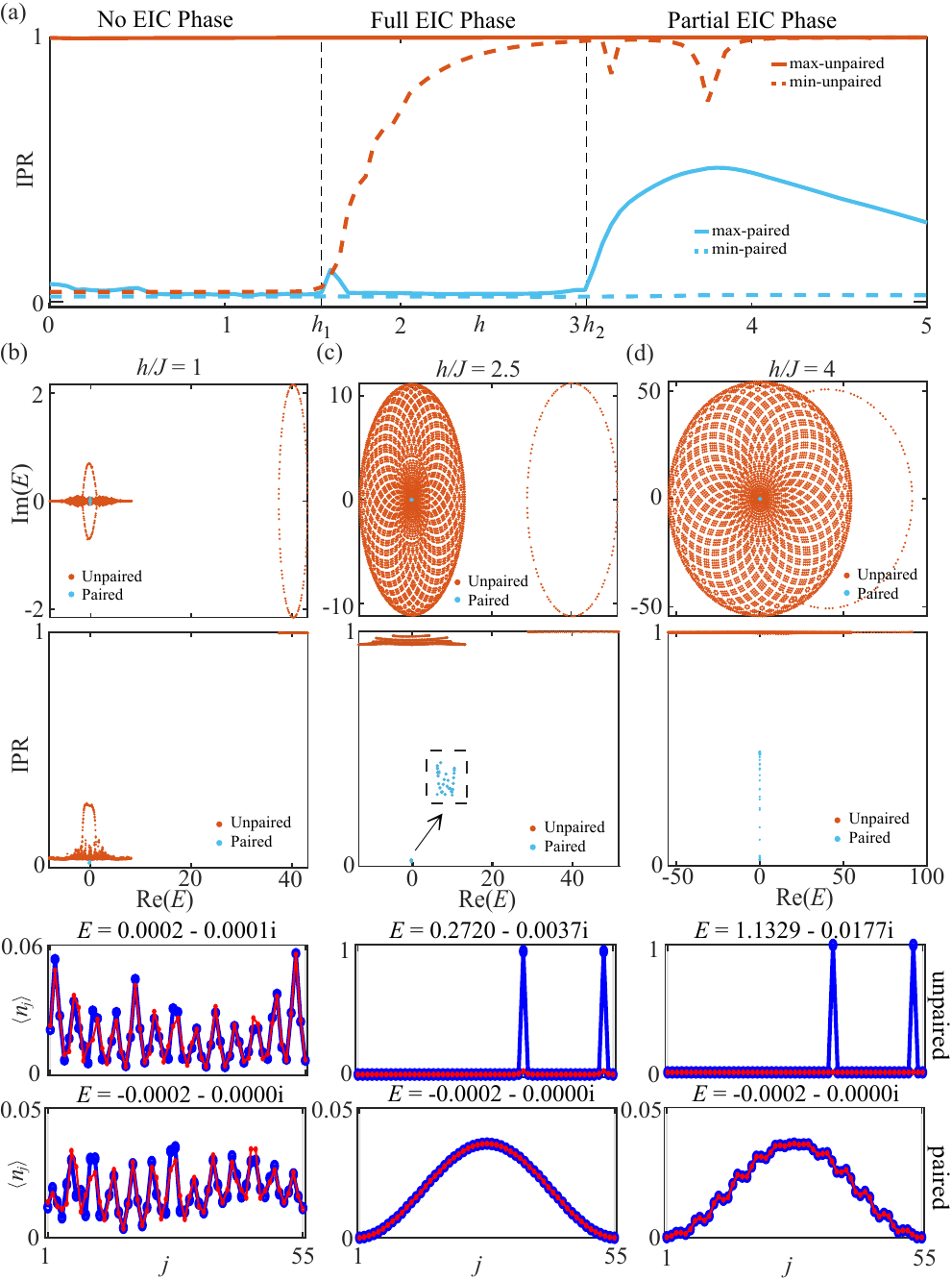}
	\caption{(a) Maximum (solid lines) and minimum (dashed lines) IPR values of the paired (light blue) and unpaired (vermilion) states as functions of the non-Hermitian strength $h$ with $U_2/J=40$. The vertical dashed lines identify the onset of the full EIC phase at $h_1\simeq1.56$ and the crossover to the partial EIC regime near $h_2\simeq3.07$. 	(b)-(d) Complex-energy spectra, IPR-resolved spectra, and representative particle-density distributions of paired and neighboring unpaired states in three characteristic regimes in (a). In the density profiles, blue and red denote the top- and bottom-leg components, respectively.
	}\label{IPR_and_spectrum_for_EIC}
\end{figure}

\section{Interaction-Stabilized Extended Composite Band in a Localized Continuum}

	We finally engineer an unconventional extended state in a localized
continuum (EIC), where an entire band of extended same-rung composite
states is embedded within localized unpaired states. Unlike the previous
case, this regime does not rely on a rung-binding interaction. Instead,
we introduce a nearest-rung interleg interaction that selectively
modifies the configuration-space pathways of the composite excitation and
stabilizes its extended dynamics.

A single intrachain hopping process transforms a same-rung configuration
$|a_jb_j\rangle$ into neighboring interleg configurations
$|a_{j+1}b_j\rangle$ and $|a_jb_{j+1}\rangle$. These configurations act
as the dominant virtual intermediate states governing the propagation and
deformation of the same-rung composite manifold. We therefore introduce
a new ladder Hamiltonian,
\begin{align}\label{H2new}
	\mathcal{H}_2
	=
	\mathcal{H}_0
	+
	\mathcal{V}(\Delta\phi=0)
	+
	U_2\sum_{j,\pm}n_{a,j}n_{b,j\pm1}.
\end{align}
The additional interaction term vanishes for same-rung configurations
$|a_jb_j\rangle$ and therefore leaves their bare energies unchanged.
Instead, it shifts the neighboring interleg configurations that enter the
virtual processes. By introducing an energy barrier in configuration
space, this interaction suppresses virtual excursions out of the
same-rung composite manifold while preserving coherent propagation of
the extended composite band.

The resulting localization behavior is summarized in
Fig.~\ref{IPR_and_spectrum_for_EIC}. For weak non-Hermiticity
($0<h<h_1$), both composite and unpaired states are extended, and no EIC
is present. Increasing $h$ drives the unpaired sector into a localized
regime, while the engineered virtual pathways allow the entire
same-rung composite band to remain extended. When the extended composite
band becomes embedded within the localized unpaired spectrum
($h_1<h<h_2$), the system realizes a full EIC phase: an extended
composite excitation residing inside a continuum of localized
unbound states. The corresponding density distributions in
Fig.~\ref{IPR_and_spectrum_for_EIC}(c) directly demonstrate that the composite band remains extended after the unpaired
continuum has become localized.

For larger $h>h_2$, the enhanced quasiperiodic modulation strengthens the
effective inhomogeneity of the composite dynamics, and the engineered
configuration-space barrier becomes insufficient to maintain the full
extended band. Some composite states become localized, leading to a
partial EIC regime.

This phenomenon is the opposite of a conventional bound state in the
continuum (BIC) \cite{PhysRevLett.133.193001,PhysRevLett.133.140202}: a BIC describes a localized
state embedded within an extended continuum, whereas the present EIC
realizes an extended composite band embedded within a localized
continuum. Further details on the biorthogonal sector weights and finite-size
properties of the embedded composite band are provided in
Appendix \ref{append4}.

\section{Conclusion}
 
 We have demonstrated configuration-selective localization and its
 controllable inversion in an interacting non-Hermitian quasiperiodic
 ladder. Although composite excitations and unbound particles occupy the
 identical microscopic lattice, their internal configurations generate
 distinct effective kinetic processes and quasiperiodic landscapes.
 Opposite leg potentials act directly on spatially separated particles
 but cancel at first order for same-rung composite states, allowing the composite band to remain extended even after the unpaired sector becomes localized. Breaking this antisymmetry restores a direct modulation of the narrow composite band and reverses the localization hierarchy,
 yielding localized composite states while the unpaired sector remains
 extended. By engineering virtual pathways in configuration space, we
 further stabilize an extended composite band embedded within a localized
 unpaired continuum. Our results establish internal configuration as an
 independent control parameter for localization and open directions for
 many-particle composites, dynamical switching, and selective transport
 and spectroscopy in synthetic lattices.

\begin{acknowledgments}
T.L. acknowledges the support from the Guangdong Provincial Quantum Science Strategic Initiative (Grant No.~GDZX2505004),  National Natural Science Foundation of China (Grant No.~12274142), the Key Program of the National Natural Science Foundation of China (Grant No.~62434009),  Introduced Innovative Team Project of Guangdong Pearl River Talents Program (Grant No.~2021ZT09Z109).
\end{acknowledgments}

\section*{Data availability}
The data that support the findings of this article are not publicly available. The data are available from the authors upon reasonable request.

\appendix

\section{Non-Hermitian Quasicrystal under Single Excitation}\label{append1}

While the main text focuses on localization inversion between composite
pairs and unbound particles in an interacting non-Hermitian
quasiperiodic ladder, we first establish the localization properties in
the single-excitation sector. This simpler limit provides a reference
for understanding how interchain coupling and quasiperiodic potentials
shape the localization landscape before interaction-induced
configuration-dependent effects emerge. It also clarifies the
transition from single-particle localization behavior to the distinct
localization properties of composite excitations in the interacting
regime.

\subsection{Single-Chain Non-Hermitian Quasicrystal}

We first consider an isolated non-Hermitian quasicrystal chain described by
\begin{equation}\label{Single}
	H_s =
	-J\sum_j
	\left(
	a_{j+1}^{\dagger}a_j+\mathrm{H.c.}
	\right)
	+
	V\sum_j
	\cos(2\pi\alpha j+ih)
	a_j^\dagger a_j .
\end{equation}

\begin{figure}[!tb]
	\centering
	\includegraphics[width=8.6cm]{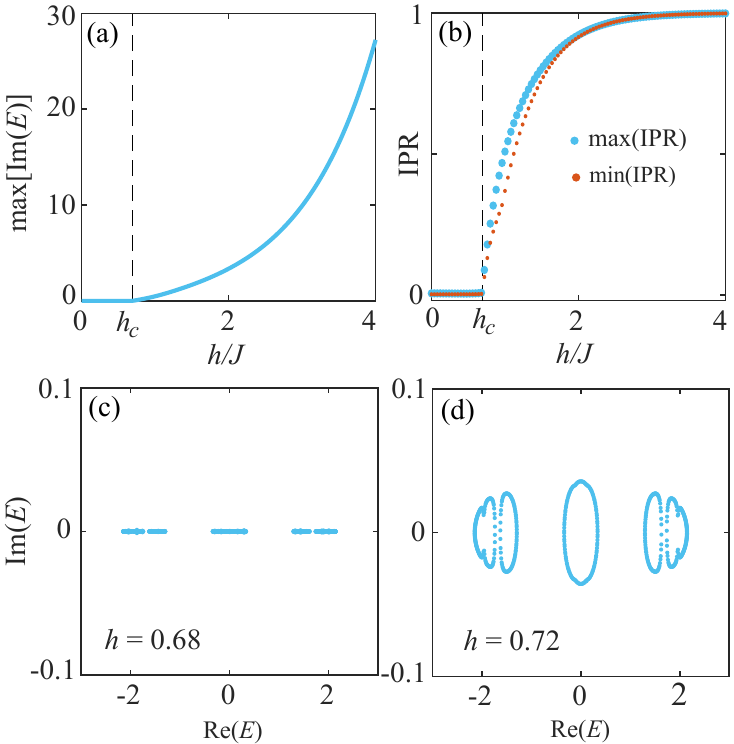}
	\caption{Localization and delocalization of single-chain non-Hermitian quasicrystal. (a) Maximum imaginary part of the complex eigenenergies as a function of the non-Hermitian strength $h$. A real-to-complex spectral transition occurs at the critical value $h/J=h_c/J\simeq0.69$. 
		(b) Minimum and maximum values of the IPR versus $h$, characterizing the localization properties of the eigenstates across the transition. 
		(c) and (d) Complex energy spectra below and above the transition, corresponding to $h/J=0.68$ and $h/J=0.72$, respectively. The parameters are $V/J=1$ and $L=610$.
	}\label{FigS1}
\end{figure}

The imaginary phase deformation $ih$ introduces non-Hermiticity into the quasiperiodic potential. For an irrational $\alpha$, this model exhibits a non-Hermitian localization transition controlled by the parameter $h$. As demonstrated in Ref.~\cite{PhysRevLett.122.237601}, the critical point is given by
\begin{equation}
	h_c=\ln(2J/V).
\end{equation}
For $h<h_c$, all eigenstates remain extended, whereas for $h>h_c$, the system enters a localized phase. This single-chain transition provides the reference point for analyzing the more complex ladder geometry, where interchain coupling and interactions generate additional localization regimes.

As shown in Fig.~\ref{FigS1}(a), the single-chain non-Hermitian quasicrystal undergoes a real-to-complex spectral transition as the non-Hermitian parameter $h$ increases. This transition occurs at the critical point $h=h_c$ and is associated with the breaking of parity-time ($\mathcal{PT}$) symmetry. Representative complex-energy spectra below and above the transition are presented in Fig.~\ref{FigS1}(c,d). Meanwhile, the inverse participation ratio (IPR) exhibits a concomitant transition from extended to localized eigenstates at the same critical point [see Fig.~\ref{FigS1}(b)], confirming the correspondence between the spectral transition and the localization transition \cite{PhysRevLett.122.237601}.

\begin{figure}[!b]
	\centering
	\includegraphics[width=8.6cm]{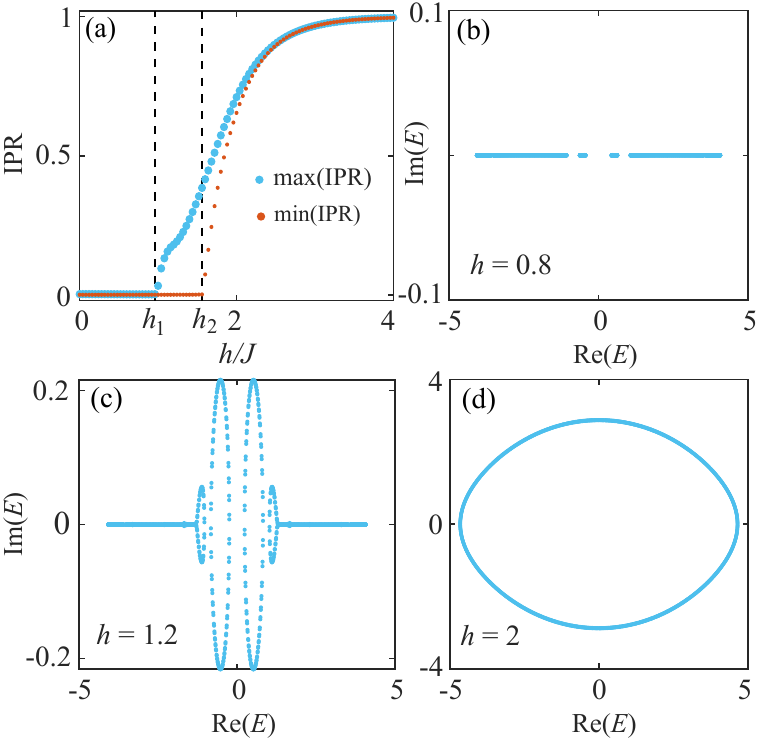}
	\caption{Localization and delocalization of   non-Hermitian quasicrystal ladder. (a) Minimum and maximum values of the IPR as a function of $h$, showing three distinct localization regimes separated by the critical points $h_1/J\simeq1$ and $h_2/J\simeq1.56$. 
		(b), (c), and (d) Complex energy spectra in the extended, mixed, and localized regimes, corresponding to $h/J=0.8$, $h/J=1.2$, and $h/J=2$, respectively. The parameters are $V/J=1$, $t/J=2$, and $L=610$.
	}\label{FigS2}
\end{figure}

\subsection{Non-Hermitian Quasicrystal Ladder with Antisymmetric Potential}

Before discussing interaction-induced localization inversion, we first
establish the localization properties of the single-excitation sector.
This provides the reference single-particle landscape from which the
configuration-dependent localization of composite excitations emerges.

The coupled non-Hermitian quasicrystal ladder with antisymmetric
potentials is described by
\begin{align}\label{eqAAH}
	H_\textrm{anti} = &  -J\sum_j \left( a_{j+1}^\dagger a_j + b_{j+1}^\dagger b_j + \text{H.c.} \right) \\ \nonumber  & - t\sum_j \left( a_j^\dagger b_j + \text{H.c.} \right) \\ \nonumber  & + V \sum_j \cos(2 \pi \alpha j   + ih) \left(  a_j^\dagger a_j-  b_j^\dagger b_j \right).	
\end{align}

The opposite quasiperiodic potentials on the two legs generate a
non-Hermitian ladder whose localization properties differ from those of
an isolated chain. As shown in Fig.~\ref{FigS2}(a), the minimum and maximum
IPR values evolve differently with increasing non-Hermitian strength
$h$, revealing three distinct regimes [see also representative complex-energy spectra in Fig.~\ref{FigS2}(b-d)]. For $h<h_1$, all single-particle
states remain extended. In the intermediate region
$h_1<h<h_2$, localized and extended states coexist, forming a
mobility-edge-like regime. For $h>h_2$, the entire single-particle
spectrum becomes localized.

The critical values $h_1$ and $h_2$ are determined not only by the
quasiperiodic modulation but also by the interchain coupling and the
antisymmetric potential structure. Therefore, the ladder geometry
provides a tunable single-particle localization landscape, which serves
as the reference background for the interaction-induced localization
inversion in the two-excitation sector.

\begin{figure}[!b]
	\centering
	\includegraphics[width=8.6cm]{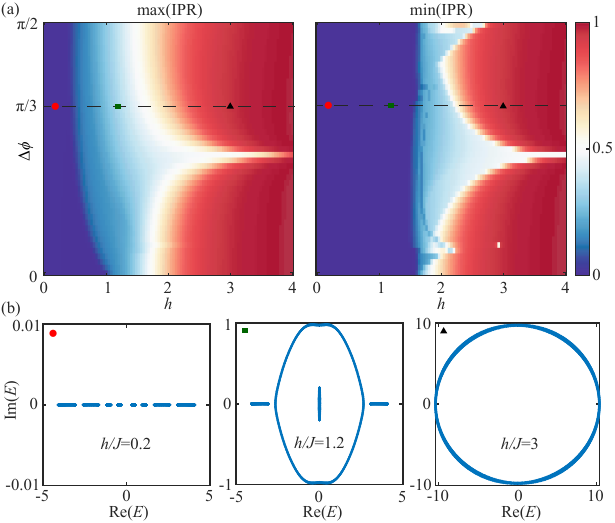}
	\caption{Localization and delocalization of   non-Hermitian quasicrystal ladder with phase difference. (a) Phase diagrams of the localization properties characterized by the IPR in the $(h,\Delta\phi)$ parameter space. The maximum (left) and minimum (right) IPR values are shown, where the common color scale indicates the degree of localization: large IPR corresponds to localized states, whereas small IPR corresponds to extended states. The phase difference $\Delta\phi$ controls the deviation from the antisymmetric quasiperiodic potential configuration. 
		(b) Complex-energy spectra for representative cases with $h/J=0.2$, $h/J=1.2$, and $h/J=1.2$ with $\Delta\phi=\pi/3$. These spectra demonstrate how breaking the potential cancellation condition modifies the localization properties of the ladder. The parameters are $V/J=1$ and $t/J=2$.
	}\label{FigS3}
\end{figure}

\subsection{Single-Excitation Localization under Broken Potential Antisymmetry}

To distinguish the interaction-induced localization inversion from
single-particle effects, we further examine how breaking the
antisymmetry of the quasiperiodic potentials modifies the single-
excitation localization landscape. Introducing a relative phase
difference $\Delta\phi$ between the two legs continuously deforms the
system away from the antisymmetric configuration and changes the
effective quasiperiodic modulation experienced by single particles.
The corresponding Hamiltonian is
\begin{align}
	\begin{aligned}
		H_\textrm{banti} = &  -J\sum_j
		\left(
		a_{j+1}^\dagger a_j
		+
		b_{j+1}^\dagger b_j
		+\mathrm{H.c.}
		\right)
		\\ \nonumber  & -t\sum_j
		\left(
		a_j^\dagger b_j+\mathrm{H.c.}
		\right)
		\\
		& +
		V\sum_j 
		\cos(2\pi\alpha j +ih+\Delta\phi/2)a_j^\dagger a_j
		\\ \nonumber  & - V\sum_j
		\cos(2\pi\alpha j +ih-\Delta\phi/2)b_j^\dagger b_j.
	\end{aligned}
	\label{eq:PD}
\end{align}

The localization phase diagram in Fig.~\ref{FigS3}(a) shows that a
finite $\Delta\phi$ shifts the localization boundaries but does not
alter the overall transition sequence. The single-excitation sector
still evolves from an extended phase to a mixed regime and eventually
to a localized phase [see also representative complex-energy spectra in Fig.~\ref{FigS3}(b)]. Therefore, the localization inversion observed
in the interacting system does not originate from a trivial change of
the single-particle localization transition, but from the distinct
effective quasiperiodic landscapes generated by particle binding and
interaction-induced virtual processes.

\section{Effective Pair Hamiltonian in the Two-Excitation Sector}\label{append2}

To understand the configuration-dependent localization behavior of bound
pairs, we derive an effective Hamiltonian for the paired subspace using
quasi-degenerate perturbation theory
\cite{Bir1974,CCohenTannoudji1Atom}. We consider the interacting
non-Hermitian quasiperiodic ladder described by
\begin{equation}
	\mathcal{H}_1
	=
	\mathcal{H}_0+\mathcal{V}(\Delta\phi)+\mathcal{H}_{\rm int},
\end{equation}
where
\begin{align}
	\mathcal{H}_0
	=
	\sum_j
	\left(
	J a_{j+1}^{\dagger}a_j
	+
	J b_{j+1}^{\dagger}b_j
	+
	t a_j^\dagger b_j
	+\mathrm{H.c.}
	\right),
\end{align}
\begin{align}
	\mathcal{V}(\Delta\phi)
	= &	V\sum_j	 \cos(2\pi\alpha j+ih+\Delta\phi/2)n_{a,j} \nonumber \\ &- V\sum_j \cos(2\pi\alpha j+ih-\Delta\phi/2)n_{b,j},
\end{align}
and
\begin{align}
	\mathcal{H}_{\rm int}
	=
	U_1\sum_j n_{a,j}n_{b,j}.
\end{align}

We focus on the strongly interacting regime
\begin{equation}
	U_1\gg |J|,|t|,|V|,
\end{equation}
where the Hilbert space separates into a paired sector and an unpaired
sector. The paired subspace is spanned by
\begin{equation}
	|\phi_j\rangle
	=
	a_j^\dagger b_j^\dagger |0\rangle ,
	\qquad j=1,\cdots,L .
\end{equation}

For the perturbative treatment, the Hamiltonian is decomposed as
\begin{equation}
	\mathcal{H}_1=H_{\rm d}+T ,
\end{equation}
where the unperturbed diagonal part contains the quasiperiodic potential
and interaction,
\begin{align}
	H_{\rm d}
	=&
	\sum_j
	\Big[
	V_{a,j}n_{a,j}
	+
	V_{b,j}n_{b,j}
	\Big]
	+
	U_1\sum_j n_{a,j}n_{b,j},
\end{align}
with
\begin{align}
	V_{a,j}
	&=
	V\cos(2\pi\alpha j+ih+\Delta\phi/2),
	\\
	V_{b,j}
	&=
	-
	V\cos(2\pi\alpha j+ih-\Delta\phi/2),
\end{align}
and \(T\) denotes the hopping terms,
\begin{align}
	T
	=
	\sum_j
	\left(
	J a_{j+1}^{\dagger}a_j
	+
	J b_{j+1}^{\dagger}b_j
	+
	t a_j^\dagger b_j
	+\mathrm{H.c.}
	\right).
\end{align}

For a same-rung pair, the zeroth-order energy is
\begin{align}
	E_{j}^{(0)}
	& = 
	U_1+V_{a,j}+V_{b,j}
	\nonumber \\ &  =
	U_1
	-
	2V\sin(2\pi\alpha j+ih)
	\sin(\Delta\phi/2).
\end{align}

For the antisymmetric case \(\Delta\phi=0\), the first-order pair
potential cancels exactly and \(E_j^{(0)}=U_1\). For finite
\(\Delta\phi\), this cancellation is broken and the pair acquires a
direct quasiperiodic modulation.

Because a single intraleg hopping event breaks the rung pair, while the
interleg hopping annihilates a same-rung pair under the hard-core
constraint,
\begin{equation}
	PTP=0 ,
\end{equation}
where
\begin{equation}
	P=\sum_j|\phi_j\rangle\langle\phi_j|,
	\qquad Q=1-P ,
\end{equation}
projects onto the paired manifold. Therefore, pair motion is generated
through virtual excursions into the unpaired sector. The exact
energy-dependent projected Hamiltonian is
\begin{equation}
	H_{\rm eff}(E)
	=
	PH_{\rm d}P+
	PTQ
	\frac{1}{E-QH_{\rm d}Q}
	QTP .
\end{equation}

For an energy-independent quasi-degenerate Hamiltonian, the second-order
matrix element is
\begin{widetext}
\begin{align}
	\langle\phi_j|
	H_{\rm eff}^{(2)}
	|\phi_{j'}\rangle
	= 
	E_j^{(0)}\delta_{jj'}
	+
	\frac{1}{2}
	\sum_{\mu\in Q}
	\langle\phi_j|T|\mu\rangle
	\langle\mu|T|\phi_{j'}\rangle
	\left[
	\frac{1}{E_j^{(0)}-E_\mu}
	+
	\frac{1}{E_{j'}^{(0)}-E_\mu}
	\right],
\end{align}
\end{widetext}
where \(|\mu\rangle\) represents an intermediate configuration with
particles on different rungs. The symmetrization of the two denominators
is required when the paired manifold is only quasi-degenerate, as occurs
for finite \(\Delta\phi\).

For pair hopping between rungs \(j\) and \(j+1\), there are two virtual
channels. Moving the particle on the upper leg gives the intermediate
state
\[
|a_{j+1}b_j\rangle ,
\]
whose energy is
\begin{equation}
	E_{a_{j+1}b_j}
	=
	V_{a,j+1}+V_{b,j}.
\end{equation}
The corresponding energy denominators associated with the initial and
final pair configurations are
\begin{align}
	E_j^{(0)}-E_{a_{j+1}b_j}
	&=
	U_1+V_{a,j}-V_{a,j+1},
	\\
	E_{j+1}^{(0)}-E_{a_{j+1}b_j}
	&=
	U_1+V_{b,j+1}-V_{b,j}.
\end{align}

Similarly, the lower-leg hopping channel through
\(|a_jb_{j+1}\rangle\) gives
\begin{align}
	E_j^{(0)}-E_{a_jb_{j+1}}
	&=
	U_1+V_{b,j}-V_{b,j+1},
	\\
	E_{j+1}^{(0)}-E_{a_jb_{j+1}}
	&=
	U_1+V_{a,j+1}-V_{a,j}.
\end{align}
Consequently, the effective pair hopping is
\begin{align}
	\mathcal{J}_j^{\rm eff}
	= &
	\frac{J^2}{2}
	\Bigg[		\frac{1}{U_1+V_{a,j}-V_{a,j+1}}
	+	\frac{1}{U_1+V_{a,j+1}-V_{a,j}}  \nonumber \\ &  +	\frac{1}{U_1+V_{b,j}-V_{b,j+1}}
	+
	\frac{1}{U_1+V_{b,j+1}-V_{b,j}}
	\Bigg].
	\label{eq:Jeff_new}
\end{align}

Virtual processes also renormalize the pair onsite energy,
\begin{align}
	\epsilon_j^{\rm eff}
	= &	\sum_{\eta=\pm1}	\left[	\frac{J^2}	{U_1+V_{a,j}-V_{a,j+\eta}}	+	\frac{J^2}
	{U_1+V_{b,j}-V_{b,j+\eta}}	\right]  \nonumber \\ &+	E_j^{(0)}.	
\end{align}\label{eq:epsilon_new}
Therefore, the effective Hamiltonian governing the composite pair becomes
\begin{align}
	H_{\rm pair}^{\rm eff}
	= &	\sum_j
	\mathcal{J}_j^{\rm eff}
	\left(
	|\phi_{j+1}\rangle\langle\phi_j|
	+
	|\phi_j\rangle\langle\phi_{j+1}|
	\right) \nonumber \\ &+	\sum_j
	\epsilon_j^{\rm eff}
	|\phi_j\rangle\langle\phi_j|.	
\end{align}\label{eq:Heffpair_new}

Equation~\eqref{eq:Heffpair_new} describes an emergent non-Hermitian
quasicrystal generated by the interaction. Importantly, the pair sector
contains two different sources of quasiperiodic modulation:

(i) a direct first-order contribution
\begin{equation}
	V_{a,j}+V_{b,j}
	=
	-2V
	\sin(2\pi\alpha j+ih)
	\sin(\Delta\phi/2),
\end{equation}
and

(ii) interaction-induced corrections generated by virtual processes,
\begin{equation}
	\mathcal{J}_j^{\rm eff},
	\epsilon_j^{\rm eff}-E_j^{(0)}
	\sim O(J^2/U_1).
\end{equation}

For \(\Delta\phi=0\), the first contribution vanishes exactly and
\(V_{b,j}=-V_{a,j}\). In this case,
Eq.~\eqref{eq:Jeff_new} reduces to
\begin{align}
	\mathcal{J}_j^{\rm eff}
	&=
	\frac{J^2}
	{U_1+V_{a,j}-V_{a,j+1}}
	+
	\frac{J^2}
	{U_1+V_{b,j}-V_{b,j+1}}
	\nonumber\\
	&=
	\frac{2J^2U_1}
	{U_1^2-
		\left(V_{a,j}-V_{a,j+1}\right)^2}.
\end{align}
For \(U_1\gg |V|\),
\begin{equation}
	\mathcal{J}_j^{\rm eff}
	\simeq
	\frac{2J^2}{U_1}
	+
	O\left(\frac{J^2V^2}{U_1^3}\right).
\end{equation}

\begin{figure}[b] 
	\centering \includegraphics[width=8.6cm]{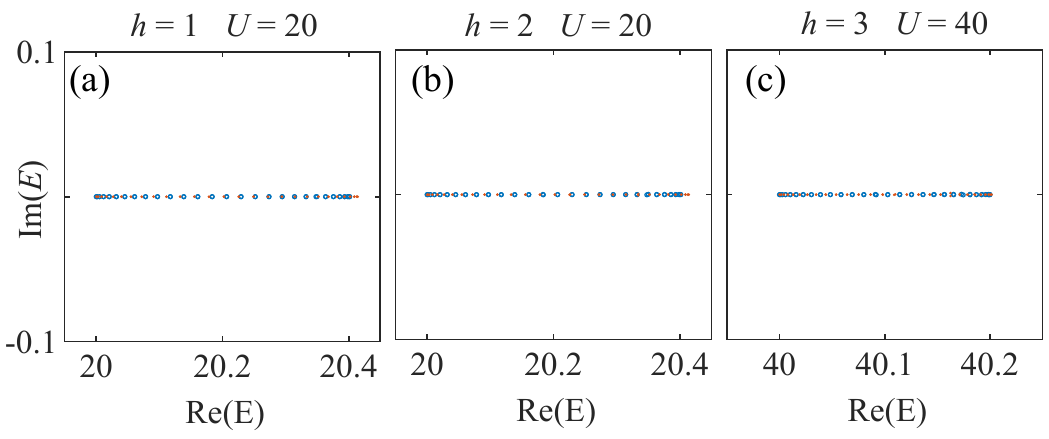} 
	\caption{ Comparison between exact diagonalization (red dots) and the effective pair Hamiltonian (blue circles) for the paired-state energy spectra. The parameters are $L=55$, $V/J=1$, with (a) $U_1/J=20$, $h=1$, (b) $U_1/J=20$, $h=2$, and (c) $U_1/J=40$, $h=3$. } \label{exact_perturbation} 
\end{figure}

Thus, the first-order quasiperiodic modulation cancels, and the remaining
spatial modulation of the effective pair Hamiltonian is generated only
through higher-order virtual corrections. The pair therefore experiences
a much weaker quasiperiodic landscape and remains extended over a broad
range of non-Hermitian strengths.

For finite \(\Delta\phi\), however, the residual first-order potential
directly acts on the narrow pair band. Because the pair bandwidth scales
as \(J^2/U_1\), this modulation can dominate the pair dynamics and induce
localization even when the corresponding unpaired particles remain
extended.

\begin{figure*}[tb]
	\centering
	\includegraphics[width=18cm]{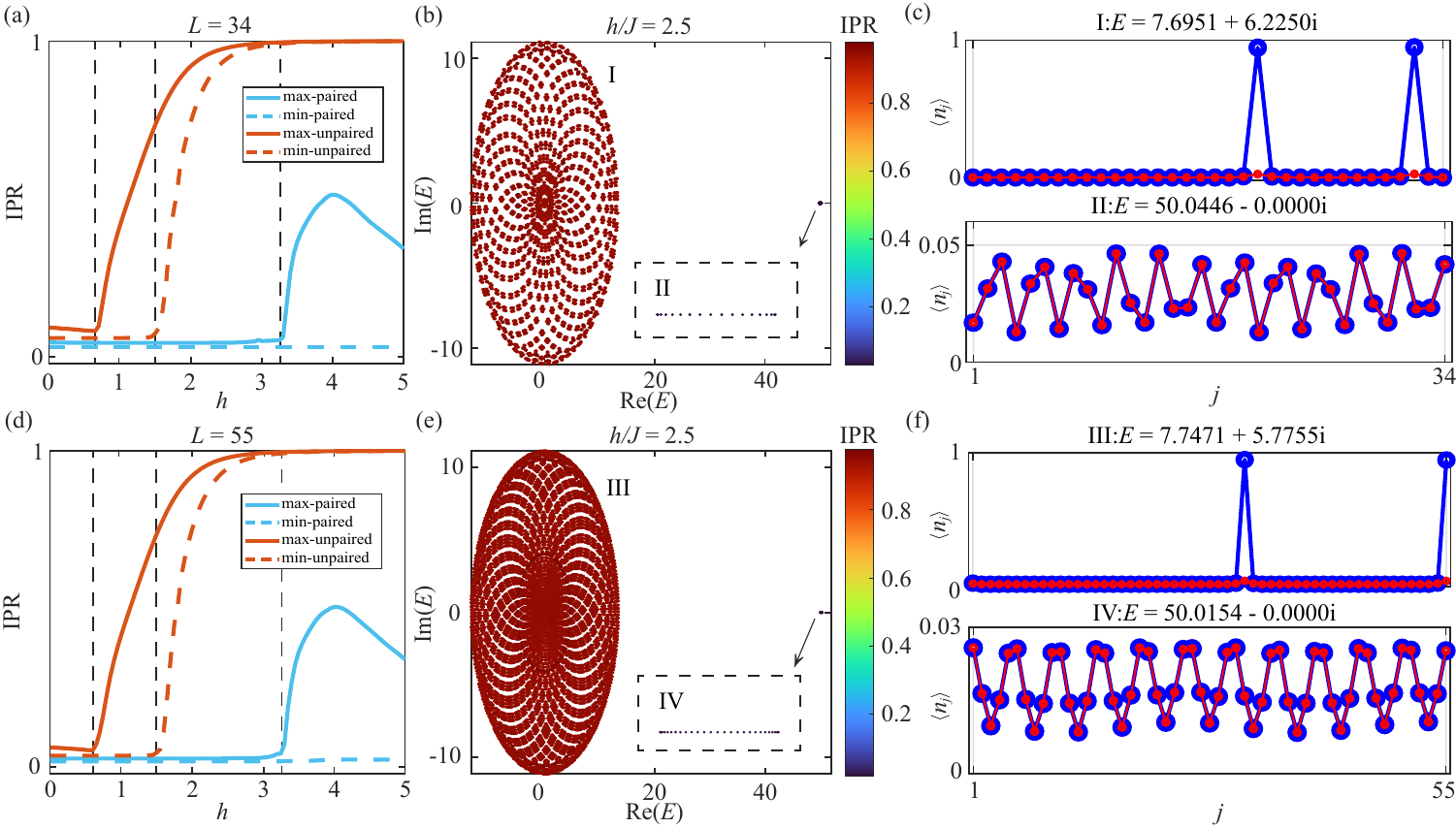}
	\caption{Finite-size robustness of the configuration-selective localization
		regime with extended composite pairs and localized unpaired states.
		(a) Maximum (solid lines) and minimum (dashed lines) 	IPR of the paired (light blue) and unpaired
		(vermilion) states as functions of the non-Hermitian strength $h$ for
		$L=34$. The vertical dashed lines indicate the characteristic
		localization boundaries of the unpaired and paired sectors.
		(b) Complex-energy spectrum at $h/J=2.5$, with the color scale denoting
		$\mathrm{IPR}$. The dashed box highlights the spectrally
		separated composite band.		(c) Rung-density profiles of a representative localized unpaired state
		(I) and an extended composite-pair state (II) selected from (b).
		(d)--(f) Corresponding results for the larger system size $L=55$.
		For both sizes, the unpaired states are localized at $h/J=2.5$, whereas
		the composite band retains a small density IPR and an extended density
		profile. The parameters are $\Delta\phi=0$, $U_1/J=50$, $V/J=1$, and
		$t/J=2$.
	}\label{fig:size_pair_localization}
\end{figure*}

We benchmark the validity of the effective Hamiltonian by comparing the energy spectra obtained from exact diagonalization of the full interacting system and from the perturbative effective model. As shown in Fig.~\ref{exact_perturbation}, the perturbative results accurately reproduce the paired-state spectra for both weak and moderate non-Hermitian strengths.

\begin{figure*}[t]
	\centering
	\includegraphics[width=18cm]{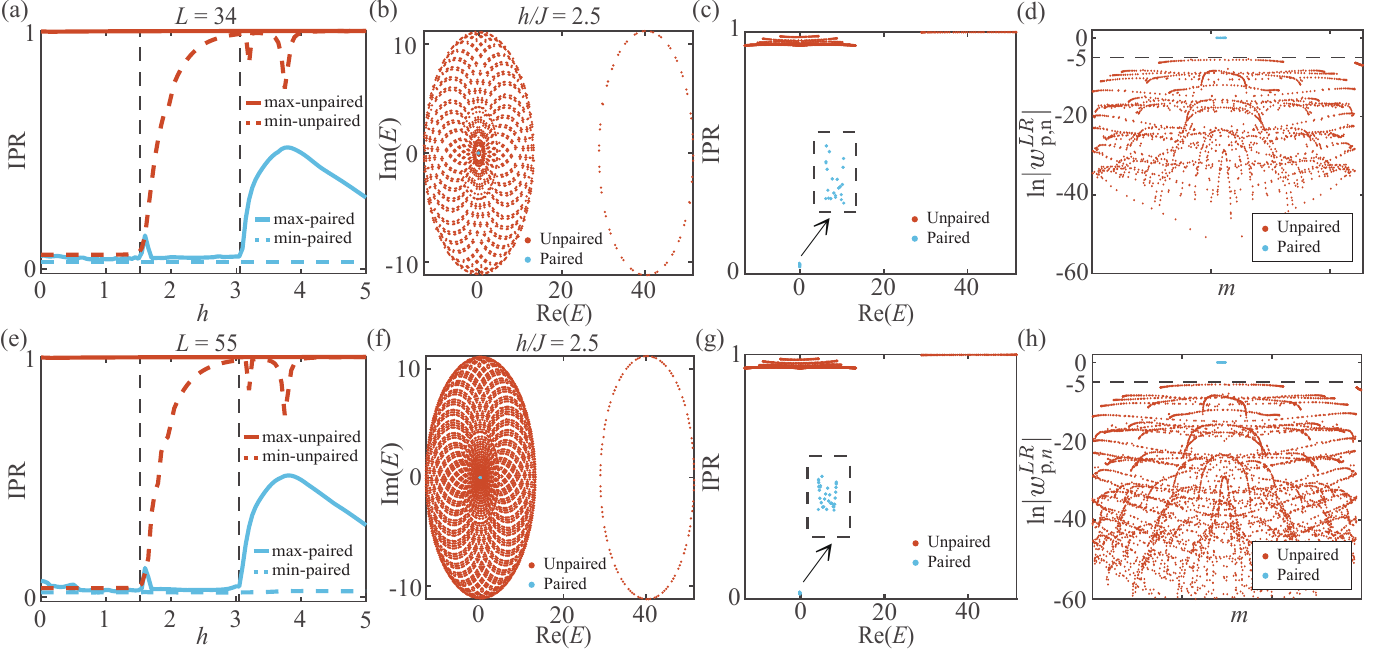}
	\caption{
		Finite-size characterization of the extended composite band embedded
		within a localized unpaired continuum.
		(a) Maximum and minimum real-space density IPR of the unpaired (vermilion) and composite
		(light blue) states as functions of $h$ for $L=34$. The vertical dashed
		lines delimit the no-EIC, full-EIC, and partial-EIC regimes.
		(b) Complex-energy spectrum at $h/J=2.5$, showing the extended
		composite band embedded within the localized unpaired spectrum.
		(c) The same spectrum resolved by $\mathrm{IPR}$, the dashed box
		highlights the embedded composite states, which retain substantially
		smaller density IPRs than the surrounding localized unpaired states.
		(d) Logarithm of the magnitude of the biorthogonal paired-sector weight,
		$\ln|w_{\mathrm p,n}^{LR}|$, for all eigenstates.
		Composite-dominated states have $|w_{\mathrm p,n}^{LR}|$ of order unity,
		whereas unpaired states have strongly suppressed paired-sector weight.
		(e)--(h) Corresponding results for $L=55$. The persistence of the
		low-IPR embedded composite band and the clear separation of the
		paired-sector weights demonstrate that the EIC regime is robust against
		increasing system size. The parameters are $h/J=2.5$ in
		(b)--(d) and (f)--(h), $U_2/J=40$, $V/J=1$, $t/J=2$, and
		$\Delta\phi=0$.
	}
	\label{fig:size_EIC}
\end{figure*}

\section{Interaction-Enhanced Configuration-Selective Localization and Finite-Size Robustness}\label{append3}

We examine how the configuration-selective localization hierarchy evolves
with the same-rung interaction $U_1$ and assess its robustness against
changes in the lattice size.
Figures~\ref{fig:size_pair_localization}(a) and
\ref{fig:size_pair_localization}(d) compare the localization evolution
for $L=34$ and $L=55$. The characteristic sequence of localization
regimes remains unchanged as the lattice size increases. In particular,
at $h/J=2.5$, the unpaired sector is already localized, whereas the
same-rung composite band remains extended. This localization contrast is
evident from both the density IPRs and the representative density
profiles.

The complex-energy spectra in
Figs.~\ref{fig:size_pair_localization}(b) and
\ref{fig:size_pair_localization}(e) show that the composite band remains
centered near $\operatorname{Re}(E)\simeq U_1$ and is spectrally
separated from most of the unpaired spectrum for both system sizes.
The representative unpaired states I and III exhibit sharply localized
density peaks, whereas the representative composite states II and IV
extend across the lattice. Upon increasing the system size from $L=34$
to $L=55$, the density IPR of the extended composite state decreases,
consistent with the expected $L^{-1}$ behavior, while the localized
unpaired-state IPR remains finite.

Compared with the $U_1/J=20$ results presented in the main text, the
larger interaction strength $U_1/J=50$ substantially broadens the
parameter interval over which the composite band remains extended after
the unpaired sector has localized. This enhancement follows from the
larger spectral separation between the paired and unpaired manifolds and
the suppression of the relative quasiperiodic modulation generated by
virtual unpaired configurations. The persistence of the same localization
hierarchy for both $L=34$ and $L=55$ confirms that the observed regime
is robust against the accessible changes in system size.

\section{Finite-size and biorthogonal characterization of the EIC	regime}\label{append4}

In this section, we  examine the finite-size robustness and configuration-space
composition of the extended composite band embedded within the localized
unpaired continuum. In contrast to the $U_1$ model, the nearest-rung
interaction $U_2$ does not shift the bare energy of a same-rung
configuration. Instead, it modifies the neighboring configurations that
mediate virtual excursions away from the same-rung manifold. This
configuration-space pathway engineering allows the composite band to
remain extended after the surrounding unpaired spectrum has become
localized.

Figures~\ref{fig:size_EIC}(a) and \ref{fig:size_EIC}(e) show the
maximum and minimum density IPRs for $L=34$ and $L=55$, respectively.
The same sequence of no-EIC, full-EIC, and partial-EIC regimes is
obtained for both lattice sizes. In the full-EIC regime, represented by
$h/J=2.5$, the unpaired states have finite density IPRs, while the
composite states retain much smaller values that decrease upon increasing
the system size. This opposite finite-size behavior confirms that the
unpaired continuum is localized whereas the embedded composite band
remains spatially extended.

The complex-energy spectra in Figs.~\ref{fig:size_EIC}(b) and
\ref{fig:size_EIC}(f) demonstrate that the composite states lie within
the spectral region occupied by the unpaired states. The IPR-resolved
spectra in Figs.~\ref{fig:size_EIC}(c) and
\ref{fig:size_EIC}(g) show that these spectrally embedded states remain
distinguishable by their small density IPRs. Importantly, the number of
unpaired states increases substantially when the lattice size is
increased, while the low-IPR composite band remains visible inside the
localized unpaired spectrum. This behavior supports the interpretation
of the composite band as an extended band embedded in a localized
continuum rather than as an accidental finite-size spectral crossing.

To identify the internal configuration of each eigenstate, we calculate
the biorthogonal paired-sector weight
\begin{equation}
	w_{\mathrm p,n}^{LR}
	=
	\frac{
		\langle L_n|P_{\mathrm p}|R_n\rangle
	}{
		\langle L_n|R_n\rangle
	},
	\qquad
	P_{\mathrm p}
	=
	\sum_{j=1}^{L}
	|\phi_j\rangle\langle\phi_j|,
\end{equation}
where $|\phi_j\rangle=a_j^\dagger b_j^\dagger|0\rangle$

Because this quantity can generally be complex, we display its magnitude
through $\ln|w_{\mathrm p,n}^{LR}|$. A composite-dominated state has
$|w_{\mathrm p,n}^{LR}|$ of order unity and therefore
$\ln|w_{\mathrm p,n}^{LR}|\simeq0$, whereas an unpaired-dominated state
has a strongly suppressed paired-sector weight and a large negative
logarithm.

As shown in Figs.~\ref{fig:size_EIC}(d) and
\ref{fig:size_EIC}(h), the states separate into a small set with
order-unity paired-sector weight and a much larger unpaired sector with
exponentially or numerically suppressed paired weight. This separation
persists when the lattice size is increased from $L=34$ to $L=55$.
Therefore, the low-IPR embedded states are not merely selected by their
spectral positions: they retain a well-defined same-rung composite
character even inside the increasingly dense localized unpaired
spectrum.

The combined density-IPR and sector-weight analyses establish two
independent properties of the EIC. The density IPR demonstrates that
the embedded composite states remain spatially extended, while
$w_{\mathrm p,n}^{LR}$ confirms that they remain dominated by the
same-rung composite manifold. Their persistence with increasing lattice
size demonstrates the robustness of the EIC mechanism against the
growth of the surrounding localized continuum.


\begin{thebibliography}{40}%
	\makeatletter
	\providecommand \@ifxundefined [1]{%
		\@ifx{#1\undefined}
	}%
	\providecommand \@ifnum [1]{%
		\ifnum #1\expandafter \@firstoftwo
		\else \expandafter \@secondoftwo
		\fi
	}%
	\providecommand \@ifx [1]{%
		\ifx #1\expandafter \@firstoftwo
		\else \expandafter \@secondoftwo
		\fi
	}%
	\providecommand \natexlab [1]{#1}%
	\providecommand \enquote  [1]{``#1''}%
	\providecommand \bibnamefont  [1]{#1}%
	\providecommand \bibfnamefont [1]{#1}%
	\providecommand \citenamefont [1]{#1}%
	\providecommand \href@noop [0]{\@secondoftwo}%
	\providecommand \href [0]{\begingroup \@sanitize@url \@href}%
	\providecommand \@href[1]{\@@startlink{#1}\@@href}%
	\providecommand \@@href[1]{\endgroup#1\@@endlink}%
	\providecommand \@sanitize@url [0]{\catcode `\\12\catcode `\$12\catcode
		`\&12\catcode `\#12\catcode `\^12\catcode `\_12\catcode `\%12\relax}%
	\providecommand \@@startlink[1]{}%
	\providecommand \@@endlink[0]{}%
	\providecommand \url  [0]{\begingroup\@sanitize@url \@url }%
	\providecommand \@url [1]{\endgroup\@href {#1}{\urlprefix }}%
	\providecommand \urlprefix  [0]{URL }%
	\providecommand \Eprint [0]{\href }%
	\providecommand \doibase [0]{http://dx.doi.org/}%
	\providecommand \selectlanguage [0]{\@gobble}%
	\providecommand \bibinfo  [0]{\@secondoftwo}%
	\providecommand \bibfield  [0]{\@secondoftwo}%
	\providecommand \translation [1]{[#1]}%
	\providecommand \BibitemOpen [0]{}%
	\providecommand \bibitemStop [0]{}%
	\providecommand \bibitemNoStop [0]{.\EOS\space}%
	\providecommand \EOS [0]{\spacefactor3000\relax}%
	\providecommand \BibitemShut  [1]{\csname bibitem#1\endcsname}%
	\let\auto@bib@innerbib\@empty
	\bibitem [{\citenamefont {Anderson}(1958)}]{PhysRev.109.1492}%
	\BibitemOpen
	\bibfield  {author} {\bibinfo {author} {\bibfnamefont {P.~W.}\ \bibnamefont
			{Anderson}},\ }\bibfield  {title} {\enquote {\bibinfo {title} {Absence of
				diffusion in certain random lattices},}\ }\href {\doibase
		10.1103/PhysRev.109.1492} {\bibfield  {journal} {\bibinfo  {journal} {Phys.
				Rev.}\ }\textbf {\bibinfo {volume} {109}},\ \bibinfo {pages} {1492} (\bibinfo
		{year} {1958})}\BibitemShut {NoStop}%
	\bibitem [{\citenamefont {Pierce}\ \emph {et~al.}(1993)\citenamefont {Pierce},
		\citenamefont {Poon},\ and\ \citenamefont {Guo}}]{Pierce1993}%
	\BibitemOpen
	\bibfield  {author} {\bibinfo {author} {\bibfnamefont {F.~S.}\ \bibnamefont
			{Pierce}}, \bibinfo {author} {\bibfnamefont {S.~J.}\ \bibnamefont {Poon}}, \
		and\ \bibinfo {author} {\bibfnamefont {Q.}~\bibnamefont {Guo}},\ }\bibfield
	{title} {\enquote {\bibinfo {title} {Electron localization in metallic
				quasicrystals},}\ }\href {\doibase 10.1126/science.261.5122.737} {\bibfield
		{journal} {\bibinfo  {journal} {Science}\ }\textbf {\bibinfo {volume}
			{261}},\ \bibinfo {pages} {737} (\bibinfo {year} {1993})}\BibitemShut
	{NoStop}%
	\bibitem [{\citenamefont {Lahini}\ \emph {et~al.}(2009)\citenamefont {Lahini},
		\citenamefont {Pugatch}, \citenamefont {Pozzi}, \citenamefont {Sorel},
		\citenamefont {Morandotti}, \citenamefont {Davidson},\ and\ \citenamefont
		{Silberberg}}]{PhysRevLett.103.013901}%
	\BibitemOpen
	\bibfield  {author} {\bibinfo {author} {\bibfnamefont {Y.}~\bibnamefont
			{Lahini}}, \bibinfo {author} {\bibfnamefont {R.}~\bibnamefont {Pugatch}},
		\bibinfo {author} {\bibfnamefont {F.}~\bibnamefont {Pozzi}}, \bibinfo
		{author} {\bibfnamefont {M.}~\bibnamefont {Sorel}}, \bibinfo {author}
		{\bibfnamefont {R.}~\bibnamefont {Morandotti}}, \bibinfo {author}
		{\bibfnamefont {N.}~\bibnamefont {Davidson}}, \ and\ \bibinfo {author}
		{\bibfnamefont {Y.}~\bibnamefont {Silberberg}},\ }\bibfield  {title}
	{\enquote {\bibinfo {title} {Observation of a localization transition in
				quasiperiodic photonic lattices},}\ }\href {\doibase
		10.1103/PhysRevLett.103.013901} {\bibfield  {journal} {\bibinfo  {journal}
			{Phys. Rev. Lett.}\ }\textbf {\bibinfo {volume} {103}},\ \bibinfo {pages}
		{013901} (\bibinfo {year} {2009})}\BibitemShut {NoStop}%
	\bibitem [{\citenamefont {Thouless}(1974)}]{Thouless1974}%
	\BibitemOpen
	\bibfield  {author} {\bibinfo {author} {\bibfnamefont {D.~J.}\ \bibnamefont
			{Thouless}},\ }\bibfield  {title} {\enquote {\bibinfo {title} {Electrons in
				disordered systems and the theory of localization},}\ }\href {\doibase
		10.1016/0370-1573(74)90029-5} {\bibfield  {journal} {\bibinfo  {journal}
			{Phys. Rep.}\ }\textbf {\bibinfo {volume} {13}},\ \bibinfo {pages} {93}
		(\bibinfo {year} {1974})}\BibitemShut {NoStop}%
	\bibitem [{\citenamefont {Winkler}\ \emph {et~al.}(2006)\citenamefont
		{Winkler}, \citenamefont {Thalhammer}, \citenamefont {Lang}, \citenamefont
		{Grimm}, \citenamefont {Hecker~Denschlag}, \citenamefont {Daley},
		\citenamefont {Kantian}, \citenamefont {B\"{u}chler},\ and\ \citenamefont
		{Zoller}}]{Winkler2006}%
	\BibitemOpen
	\bibfield  {author} {\bibinfo {author} {\bibfnamefont {K.}~\bibnamefont
			{Winkler}}, \bibinfo {author} {\bibfnamefont {G.}~\bibnamefont {Thalhammer}},
		\bibinfo {author} {\bibfnamefont {F.}~\bibnamefont {Lang}}, \bibinfo {author}
		{\bibfnamefont {R.}~\bibnamefont {Grimm}}, \bibinfo {author} {\bibfnamefont
			{J.}~\bibnamefont {Hecker~Denschlag}}, \bibinfo {author} {\bibfnamefont
			{A.~J.}\ \bibnamefont {Daley}}, \bibinfo {author} {\bibfnamefont
			{A.}~\bibnamefont {Kantian}}, \bibinfo {author} {\bibfnamefont {H.~P.}\
			\bibnamefont {B\"{u}chler}}, \ and\ \bibinfo {author} {\bibfnamefont
			{P.}~\bibnamefont {Zoller}},\ }\bibfield  {title} {\enquote {\bibinfo {title}
			{Repulsively bound atom pairs in an optical lattice},}\ }\href {\doibase
		10.1038/nature04918} {\bibfield  {journal} {\bibinfo  {journal} {Nature}\
		}\textbf {\bibinfo {volume} {441}},\ \bibinfo {pages} {853} (\bibinfo {year}
		{2006})}\BibitemShut {NoStop}%
	\bibitem [{\citenamefont {Shepelyansky}(1994)}]{PhysRevLett.73.2607}%
	\BibitemOpen
	\bibfield  {author} {\bibinfo {author} {\bibfnamefont {D.~L.}\ \bibnamefont
			{Shepelyansky}},\ }\bibfield  {title} {\enquote {\bibinfo {title} {Coherent
				propagation of two interacting particles in a random potential},}\ }\href
	{\doibase 10.1103/PhysRevLett.73.2607} {\bibfield  {journal} {\bibinfo
			{journal} {Phys. Rev. Lett.}\ }\textbf {\bibinfo {volume} {73}},\ \bibinfo
		{pages} {2607} (\bibinfo {year} {1994})}\BibitemShut {NoStop}%
	\bibitem [{\citenamefont {Levine}\ and\ \citenamefont
		{Steinhardt}(1984)}]{PhysRevLett.53.2477}%
	\BibitemOpen
	\bibfield  {author} {\bibinfo {author} {\bibfnamefont {D.}~\bibnamefont
			{Levine}}\ and\ \bibinfo {author} {\bibfnamefont {Paul~J.}\ \bibnamefont
			{Steinhardt}},\ }\bibfield  {title} {\enquote {\bibinfo {title}
			{Quasicrystals: {A} new class of ordered structures},}\ }\href {\doibase
		10.1103/PhysRevLett.53.2477} {\bibfield  {journal} {\bibinfo  {journal}
			{Phys. Rev. Lett.}\ }\textbf {\bibinfo {volume} {53}},\ \bibinfo {pages}
		{2477} (\bibinfo {year} {1984})}\BibitemShut {NoStop}%
	\bibitem [{\citenamefont {Goblot}\ \emph {et~al.}(2020)\citenamefont {Goblot},
		\citenamefont {Štrkalj}, \citenamefont {Pernet}, \citenamefont {Lado},
		\citenamefont {Dorow}, \citenamefont {Lemaître}, \citenamefont {Le~Gratiet},
		\citenamefont {Harouri}, \citenamefont {Sagnes}, \citenamefont {Ravets},
		\citenamefont {Amo}, \citenamefont {Bloch},\ and\ \citenamefont
		{Zilberberg}}]{Goblot2020}%
	\BibitemOpen
	\bibfield  {author} {\bibinfo {author} {\bibfnamefont {V.}~\bibnamefont
			{Goblot}}, \bibinfo {author} {\bibfnamefont {A.}~\bibnamefont {Štrkalj}},
		\bibinfo {author} {\bibfnamefont {N.}~\bibnamefont {Pernet}}, \bibinfo
		{author} {\bibfnamefont {J.~L.}\ \bibnamefont {Lado}}, \bibinfo {author}
		{\bibfnamefont {C.}~\bibnamefont {Dorow}}, \bibinfo {author} {\bibfnamefont
			{A.}~\bibnamefont {Lemaître}}, \bibinfo {author} {\bibfnamefont
			{L.}~\bibnamefont {Le~Gratiet}}, \bibinfo {author} {\bibfnamefont
			{A.}~\bibnamefont {Harouri}}, \bibinfo {author} {\bibfnamefont
			{I.}~\bibnamefont {Sagnes}}, \bibinfo {author} {\bibfnamefont
			{S.}~\bibnamefont {Ravets}}, \bibinfo {author} {\bibfnamefont
			{A.}~\bibnamefont {Amo}}, \bibinfo {author} {\bibfnamefont {J.}~\bibnamefont
			{Bloch}}, \ and\ \bibinfo {author} {\bibfnamefont {O.}~\bibnamefont
			{Zilberberg}},\ }\bibfield  {title} {\enquote {\bibinfo {title} {Emergence of
				criticality through a cascade of delocalization transitions in quasiperiodic
				chains},}\ }\href {\doibase 10.1038/s41567-020-0908-7} {\bibfield  {journal}
		{\bibinfo  {journal} {Nat. Phys.}\ }\textbf {\bibinfo {volume} {16}},\
		\bibinfo {pages} {832} (\bibinfo {year} {2020})}\BibitemShut {NoStop}%
	\bibitem [{\citenamefont {Wang}\ \emph {et~al.}(2020)\citenamefont {Wang},
		\citenamefont {Xia}, \citenamefont {Zhang}, \citenamefont {Yao},
		\citenamefont {Chen}, \citenamefont {You}, \citenamefont {Zhou},\ and\
		\citenamefont {Liu}}]{PhysRevLett.125.196604}%
	\BibitemOpen
	\bibfield  {author} {\bibinfo {author} {\bibfnamefont {Y.}~\bibnamefont
			{Wang}}, \bibinfo {author} {\bibfnamefont {X.}~\bibnamefont {Xia}}, \bibinfo
		{author} {\bibfnamefont {L.}~\bibnamefont {Zhang}}, \bibinfo {author}
		{\bibfnamefont {H.}~\bibnamefont {Yao}}, \bibinfo {author} {\bibfnamefont
			{S.}~\bibnamefont {Chen}}, \bibinfo {author} {\bibfnamefont {J.}~\bibnamefont
			{You}}, \bibinfo {author} {\bibfnamefont {Q.}~\bibnamefont {Zhou}}, \ and\
		\bibinfo {author} {\bibfnamefont {X.-J.}\ \bibnamefont {Liu}},\ }\bibfield
	{title} {\enquote {\bibinfo {title} {One-dimensional quasiperiodic mosaic
				lattice with exact mobility edges},}\ }\href {\doibase
		10.1103/PhysRevLett.125.196604} {\bibfield  {journal} {\bibinfo  {journal}
			{Phys. Rev. Lett.}\ }\textbf {\bibinfo {volume} {125}},\ \bibinfo {pages}
		{196604} (\bibinfo {year} {2020})}\BibitemShut {NoStop}%
	\bibitem [{\citenamefont {Wang}\ \emph {et~al.}(2022)\citenamefont {Wang},
		\citenamefont {Zhang}, \citenamefont {Sun}, \citenamefont {Poon},\ and\
		\citenamefont {Liu}}]{PhysRevB.106.L140203}%
	\BibitemOpen
	\bibfield  {author} {\bibinfo {author} {\bibfnamefont {Y.}~\bibnamefont
			{Wang}}, \bibinfo {author} {\bibfnamefont {L.}~\bibnamefont {Zhang}},
		\bibinfo {author} {\bibfnamefont {W.}~\bibnamefont {Sun}}, \bibinfo {author}
		{\bibfnamefont {T.-F.~J.}\ \bibnamefont {Poon}}, \ and\ \bibinfo {author}
		{\bibfnamefont {X.-J.}\ \bibnamefont {Liu}},\ }\bibfield  {title} {\enquote
		{\bibinfo {title} {Quantum phase with coexisting localized, extended, and
				critical zones},}\ }\href {\doibase 10.1103/PhysRevB.106.L140203} {\bibfield
		{journal} {\bibinfo  {journal} {Phys. Rev. B}\ }\textbf {\bibinfo {volume}
			{106}},\ \bibinfo {pages} {L140203} (\bibinfo {year} {2022})}\BibitemShut
	{NoStop}%
	\bibitem [{\citenamefont {Wang}\ \emph {et~al.}(2024)\citenamefont {Wang},
		\citenamefont {Fu}, \citenamefont {Konotop}, \citenamefont {Kartashov},\ and\
		\citenamefont {Ye}}]{Wang2024}%
	\BibitemOpen
	\bibfield  {author} {\bibinfo {author} {\bibfnamefont {P.}~\bibnamefont
			{Wang}}, \bibinfo {author} {\bibfnamefont {Q.}~\bibnamefont {Fu}}, \bibinfo
		{author} {\bibfnamefont {V.~V.}\ \bibnamefont {Konotop}}, \bibinfo {author}
		{\bibfnamefont {Y.~V.}\ \bibnamefont {Kartashov}}, \ and\ \bibinfo {author}
		{\bibfnamefont {F.}~\bibnamefont {Ye}},\ }\bibfield  {title} {\enquote
		{\bibinfo {title} {Observation of localization of light in linear photonic
				quasicrystals with diverse rotational symmetries},}\ }\href {\doibase
		10.1038/s41566-023-01350-6} {\bibfield  {journal} {\bibinfo  {journal} {Nat.
				Photon.}\ }\textbf {\bibinfo {volume} {18}},\ \bibinfo {pages} {224}
		(\bibinfo {year} {2024})}\BibitemShut {NoStop}%
	\bibitem [{\citenamefont {Yao}\ and\ \citenamefont
		{Wang}(2018)}]{ShunyuYao2018}%
	\BibitemOpen
	\bibfield  {author} {\bibinfo {author} {\bibfnamefont {S.}~\bibnamefont
			{Yao}}\ and\ \bibinfo {author} {\bibfnamefont {Z.}~\bibnamefont {Wang}},\
	}\bibfield  {title} {\enquote {\bibinfo {title} {Edge states and topological
				invariants of non-\uppercase{H}ermitian systems},}\ }\href
	{https://link.aps.org/doi/10.1103/PhysRevLett.121.086803} {\bibfield
		{journal} {\bibinfo  {journal} {Phys. Rev. Lett.}\ }\textbf {\bibinfo
			{volume} {121}},\ \bibinfo {pages} {086803} (\bibinfo {year}
		{2018})}\BibitemShut {NoStop}%
	\bibitem [{\citenamefont {Yokomizo}\ and\ \citenamefont
		{Murakami}(2019)}]{PhysRevLett.123.066404}%
	\BibitemOpen
	\bibfield  {author} {\bibinfo {author} {\bibfnamefont {K.}~\bibnamefont
			{Yokomizo}}\ and\ \bibinfo {author} {\bibfnamefont {S.}~\bibnamefont
			{Murakami}},\ }\bibfield  {title} {\enquote {\bibinfo {title} {Non-{B}loch
				band theory of non-{H}ermitian systems},}\ }\href {\doibase
		10.1103/PhysRevLett.123.066404} {\bibfield  {journal} {\bibinfo  {journal}
			{Phys. Rev. Lett.}\ }\textbf {\bibinfo {volume} {123}},\ \bibinfo {pages}
		{066404} (\bibinfo {year} {2019})}\BibitemShut {NoStop}%
	\bibitem [{\citenamefont {Zhang}\ \emph {et~al.}(2020)\citenamefont {Zhang},
		\citenamefont {Yang},\ and\ \citenamefont {Fang}}]{PhysRevLett.125.126402}%
	\BibitemOpen
	\bibfield  {author} {\bibinfo {author} {\bibfnamefont {K.}~\bibnamefont
			{Zhang}}, \bibinfo {author} {\bibfnamefont {Z.}~\bibnamefont {Yang}}, \ and\
		\bibinfo {author} {\bibfnamefont {C.}~\bibnamefont {Fang}},\ }\bibfield
	{title} {\enquote {\bibinfo {title} {Correspondence between winding numbers
				and skin modes in non-{H}ermitian systems},}\ }\href {\doibase
		10.1103/PhysRevLett.125.126402} {\bibfield  {journal} {\bibinfo  {journal}
			{Phys. Rev. Lett.}\ }\textbf {\bibinfo {volume} {125}},\ \bibinfo {pages}
		{126402} (\bibinfo {year} {2020})}\BibitemShut {NoStop}%
	\bibitem [{\citenamefont {Liu}\ \emph {et~al.}(2019)\citenamefont {Liu},
		\citenamefont {Zhang}, \citenamefont {Ai}, \citenamefont {Gong},
		\citenamefont {Kawabata}, \citenamefont {Ueda},\ and\ \citenamefont
		{Nori}}]{PhysRevLett.122.076801}%
	\BibitemOpen
	\bibfield  {author} {\bibinfo {author} {\bibfnamefont {T.}~\bibnamefont
			{Liu}}, \bibinfo {author} {\bibfnamefont {Y.-R.}\ \bibnamefont {Zhang}},
		\bibinfo {author} {\bibfnamefont {Q.}~\bibnamefont {Ai}}, \bibinfo {author}
		{\bibfnamefont {Z.}~\bibnamefont {Gong}}, \bibinfo {author} {\bibfnamefont
			{K.}~\bibnamefont {Kawabata}}, \bibinfo {author} {\bibfnamefont
			{M.}~\bibnamefont {Ueda}}, \ and\ \bibinfo {author} {\bibfnamefont
			{F.}~\bibnamefont {Nori}},\ }\bibfield  {title} {\enquote {\bibinfo {title}
			{Second-order topological phases in non-{H}ermitian systems},}\ }\href
	{\doibase 10.1103/PhysRevLett.122.076801} {\bibfield  {journal} {\bibinfo
			{journal} {Phys. Rev. Lett.}\ }\textbf {\bibinfo {volume} {122}},\ \bibinfo
		{pages} {076801} (\bibinfo {year} {2019})}\BibitemShut {NoStop}%
	\bibitem [{\citenamefont {Kunst}\ \emph {et~al.}(2018)\citenamefont {Kunst},
		\citenamefont {Edvardsson}, \citenamefont {Budich},\ and\ \citenamefont
		{Bergholtz}}]{PhysRevLett.121.026808}%
	\BibitemOpen
	\bibfield  {author} {\bibinfo {author} {\bibfnamefont {F.~K.}\ \bibnamefont
			{Kunst}}, \bibinfo {author} {\bibfnamefont {E.}~\bibnamefont {Edvardsson}},
		\bibinfo {author} {\bibfnamefont {J.~C.}\ \bibnamefont {Budich}}, \ and\
		\bibinfo {author} {\bibfnamefont {E.~J.}\ \bibnamefont {Bergholtz}},\
	}\bibfield  {title} {\enquote {\bibinfo {title} {Biorthogonal bulk-boundary
				correspondence in non-{H}ermitian systems},}\ }\href {\doibase
		10.1103/PhysRevLett.121.026808} {\bibfield  {journal} {\bibinfo  {journal}
			{Phys. Rev. Lett.}\ }\textbf {\bibinfo {volume} {121}},\ \bibinfo {pages}
		{026808} (\bibinfo {year} {2018})}\BibitemShut {NoStop}%
	\bibitem [{\citenamefont {Leefmans}\ \emph {et~al.}(2022)\citenamefont
		{Leefmans}, \citenamefont {Dutt}, \citenamefont {Williams}, \citenamefont
		{Yuan}, \citenamefont {Parto}, \citenamefont {Nori}, \citenamefont {Fan},\
		and\ \citenamefont {Marandi}}]{Leefmans2022}%
	\BibitemOpen
	\bibfield  {author} {\bibinfo {author} {\bibfnamefont {C.}~\bibnamefont
			{Leefmans}}, \bibinfo {author} {\bibfnamefont {A.}~\bibnamefont {Dutt}},
		\bibinfo {author} {\bibfnamefont {J.}~\bibnamefont {Williams}}, \bibinfo
		{author} {\bibfnamefont {L.}~\bibnamefont {Yuan}}, \bibinfo {author}
		{\bibfnamefont {M.}~\bibnamefont {Parto}}, \bibinfo {author} {\bibfnamefont
			{F.}~\bibnamefont {Nori}}, \bibinfo {author} {\bibfnamefont {S.}~\bibnamefont
			{Fan}}, \ and\ \bibinfo {author} {\bibfnamefont {A.}~\bibnamefont
			{Marandi}},\ }\bibfield  {title} {\enquote {\bibinfo {title} {Topological
				dissipation in a time-multiplexed photonic resonator network},}\ }\href
	{\doibase 10.1038/s41567-021-01492-w} {\bibfield  {journal} {\bibinfo
			{journal} {Nat. Phys.}\ }\textbf {\bibinfo {volume} {18}},\ \bibinfo {pages}
		{442} (\bibinfo {year} {2022})}\BibitemShut {NoStop}%
	\bibitem [{\citenamefont {Gong}\ \emph {et~al.}(2018)\citenamefont {Gong},
		\citenamefont {Ashida}, \citenamefont {Kawabata}, \citenamefont {Takasan},
		\citenamefont {Higashikawa},\ and\ \citenamefont {Ueda}}]{arXiv:1802.07964}%
	\BibitemOpen
	\bibfield  {author} {\bibinfo {author} {\bibfnamefont {Z.}~\bibnamefont
			{Gong}}, \bibinfo {author} {\bibfnamefont {Y.}~\bibnamefont {Ashida}},
		\bibinfo {author} {\bibfnamefont {K.}~\bibnamefont {Kawabata}}, \bibinfo
		{author} {\bibfnamefont {K.}~\bibnamefont {Takasan}}, \bibinfo {author}
		{\bibfnamefont {S.}~\bibnamefont {Higashikawa}}, \ and\ \bibinfo {author}
		{\bibfnamefont {M.}~\bibnamefont {Ueda}},\ }\bibfield  {title} {\enquote
		{\bibinfo {title} {Topological phases of non-\uppercase{H}ermitian
				systems},}\ }\href {https://link.aps.org/doi/10.1103/PhysRevX.8.031079}
	{\bibfield  {journal} {\bibinfo  {journal} {Phys. Rev. X}\ }\textbf {\bibinfo
			{volume} {8}},\ \bibinfo {pages} {031079} (\bibinfo {year}
		{2018})}\BibitemShut {NoStop}%
	\bibitem [{\citenamefont {Lee}\ \emph {et~al.}(2019{\natexlab{a}})\citenamefont
		{Lee}, \citenamefont {Li},\ and\ \citenamefont
		{Gong}}]{PhysRevLett.123.016805}%
	\BibitemOpen
	\bibfield  {author} {\bibinfo {author} {\bibfnamefont {C.~H.}\ \bibnamefont
			{Lee}}, \bibinfo {author} {\bibfnamefont {L.}~\bibnamefont {Li}}, \ and\
		\bibinfo {author} {\bibfnamefont {J.}~\bibnamefont {Gong}},\ }\bibfield
	{title} {\enquote {\bibinfo {title} {Hybrid higher-order skin-topological
				modes in nonreciprocal systems},}\ }\href {\doibase
		10.1103/PhysRevLett.123.016805} {\bibfield  {journal} {\bibinfo  {journal}
			{Phys. Rev. Lett.}\ }\textbf {\bibinfo {volume} {123}},\ \bibinfo {pages}
		{016805} (\bibinfo {year} {2019}{\natexlab{a}})}\BibitemShut {NoStop}%
	\bibitem [{\citenamefont {Lee}\ \emph {et~al.}(2019{\natexlab{b}})\citenamefont
		{Lee}, \citenamefont {Ahn}, \citenamefont {Zhou},\ and\ \citenamefont
		{Vishwanath}}]{PhysRevLett.123.206404}%
	\BibitemOpen
	\bibfield  {author} {\bibinfo {author} {\bibfnamefont {J.~Y.}\ \bibnamefont
			{Lee}}, \bibinfo {author} {\bibfnamefont {J.}~\bibnamefont {Ahn}}, \bibinfo
		{author} {\bibfnamefont {H.}~\bibnamefont {Zhou}}, \ and\ \bibinfo {author}
		{\bibfnamefont {A.}~\bibnamefont {Vishwanath}},\ }\bibfield  {title}
	{\enquote {\bibinfo {title} {Topological correspondence between {H}ermitian
				and non-{H}ermitian systems: {A}nomalous dynamics},}\ }\href {\doibase
		10.1103/PhysRevLett.123.206404} {\bibfield  {journal} {\bibinfo  {journal}
			{Phys. Rev. Lett.}\ }\textbf {\bibinfo {volume} {123}},\ \bibinfo {pages}
		{206404} (\bibinfo {year} {2019}{\natexlab{b}})}\BibitemShut {NoStop}%
	\bibitem [{\citenamefont {Kawabata}\ \emph {et~al.}(2019)\citenamefont
		{Kawabata}, \citenamefont {Shiozaki}, \citenamefont {Ueda},\ and\
		\citenamefont {Sato}}]{PhysRevX.9.041015}%
	\BibitemOpen
	\bibfield  {author} {\bibinfo {author} {\bibfnamefont {K.}~\bibnamefont
			{Kawabata}}, \bibinfo {author} {\bibfnamefont {K.}~\bibnamefont {Shiozaki}},
		\bibinfo {author} {\bibfnamefont {M.}~\bibnamefont {Ueda}}, \ and\ \bibinfo
		{author} {\bibfnamefont {M.}~\bibnamefont {Sato}},\ }\bibfield  {title}
	{\enquote {\bibinfo {title} {Symmetry and topology in non-{H}ermitian
				physics},}\ }\href {\doibase 10.1103/PhysRevX.9.041015} {\bibfield  {journal}
		{\bibinfo  {journal} {Phys. Rev. X}\ }\textbf {\bibinfo {volume} {9}},\
		\bibinfo {pages} {041015} (\bibinfo {year} {2019})}\BibitemShut {NoStop}%
	\bibitem [{\citenamefont {Okuma}\ \emph {et~al.}(2020)\citenamefont {Okuma},
		\citenamefont {Kawabata}, \citenamefont {Shiozaki},\ and\ \citenamefont
		{Sato}}]{PhysRevLett.124.086801}%
	\BibitemOpen
	\bibfield  {author} {\bibinfo {author} {\bibfnamefont {N.}~\bibnamefont
			{Okuma}}, \bibinfo {author} {\bibfnamefont {K.}~\bibnamefont {Kawabata}},
		\bibinfo {author} {\bibfnamefont {K.}~\bibnamefont {Shiozaki}}, \ and\
		\bibinfo {author} {\bibfnamefont {M.}~\bibnamefont {Sato}},\ }\bibfield
	{title} {\enquote {\bibinfo {title} {Topological origin of non-{H}ermitian
				skin effects},}\ }\href {\doibase 10.1103/PhysRevLett.124.086801} {\bibfield
		{journal} {\bibinfo  {journal} {Phys. Rev. Lett.}\ }\textbf {\bibinfo
			{volume} {124}},\ \bibinfo {pages} {086801} (\bibinfo {year}
		{2020})}\BibitemShut {NoStop}%
	\bibitem [{\citenamefont {Liu}\ \emph {et~al.}(2021)\citenamefont {Liu},
		\citenamefont {He}, \citenamefont {Yang},\ and\ \citenamefont
		{Nori}}]{PhysRevLett.127.196801}%
	\BibitemOpen
	\bibfield  {author} {\bibinfo {author} {\bibfnamefont {T.}~\bibnamefont
			{Liu}}, \bibinfo {author} {\bibfnamefont {J.~J.}\ \bibnamefont {He}},
		\bibinfo {author} {\bibfnamefont {Z.}~\bibnamefont {Yang}}, \ and\ \bibinfo
		{author} {\bibfnamefont {F.}~\bibnamefont {Nori}},\ }\bibfield  {title}
	{\enquote {\bibinfo {title} {Higher-order {W}eyl-exceptional-ring
				semimetals},}\ }\href {\doibase 10.1103/PhysRevLett.127.196801} {\bibfield
		{journal} {\bibinfo  {journal} {Phys. Rev. Lett.}\ }\textbf {\bibinfo
			{volume} {127}},\ \bibinfo {pages} {196801} (\bibinfo {year}
		{2021})}\BibitemShut {NoStop}%
	\bibitem [{\citenamefont {Cai}\ \emph {et~al.}(2026)\citenamefont {Cai},
		\citenamefont {Li}, \citenamefont {Zhang}, \citenamefont {Wei}, \citenamefont
		{Yang}, \citenamefont {Liu},\ and\ \citenamefont {Nori}}]{zf4k-ytgt}%
	\BibitemOpen
	\bibfield  {author} {\bibinfo {author} {\bibfnamefont {Z.-F.}\ \bibnamefont
			{Cai}}, \bibinfo {author} {\bibfnamefont {Y.}~\bibnamefont {Li}}, \bibinfo
		{author} {\bibfnamefont {Y.-R.}\ \bibnamefont {Zhang}}, \bibinfo {author}
		{\bibfnamefont {X.}~\bibnamefont {Wei}}, \bibinfo {author} {\bibfnamefont
			{Z.}~\bibnamefont {Yang}}, \bibinfo {author} {\bibfnamefont {T.}~\bibnamefont
			{Liu}}, \ and\ \bibinfo {author} {\bibfnamefont {F.}~\bibnamefont {Nori}},\
	}\bibfield  {title} {\enquote {\bibinfo {title} {Arbitrary control of
				non-{H}ermitian skin modes via disorder and an electric field},}\ }\href
	{\doibase 10.1103/zf4k-ytgt} {\bibfield  {journal} {\bibinfo  {journal}
			{Phys. Rev. Res.}\ }\textbf {\bibinfo {volume} {8}},\ \bibinfo {pages}
		{033021} (\bibinfo {year} {2026})}\BibitemShut {NoStop}%
	\bibitem [{\citenamefont {Li}\ and\ \citenamefont
		{Xu}(2022)}]{PhysRevLett.129.093001}%
	\BibitemOpen
	\bibfield  {author} {\bibinfo {author} {\bibfnamefont {K.}~\bibnamefont
			{Li}}\ and\ \bibinfo {author} {\bibfnamefont {Y.}~\bibnamefont {Xu}},\
	}\bibfield  {title} {\enquote {\bibinfo {title} {Non-{H}ermitian absorption
				spectroscopy},}\ }\href {\doibase 10.1103/PhysRevLett.129.093001} {\bibfield
		{journal} {\bibinfo  {journal} {Phys. Rev. Lett.}\ }\textbf {\bibinfo
			{volume} {129}},\ \bibinfo {pages} {093001} (\bibinfo {year}
		{2022})}\BibitemShut {NoStop}%
	\bibitem [{\citenamefont {Kawabata}\ and\ \citenamefont
		{Nakamura}(2025)}]{vxgf-59xt}%
	\BibitemOpen
	\bibfield  {author} {\bibinfo {author} {\bibfnamefont {K.}~\bibnamefont
			{Kawabata}}\ and\ \bibinfo {author} {\bibfnamefont {D.}~\bibnamefont
			{Nakamura}},\ }\bibfield  {title} {\enquote {\bibinfo {title} {Hopf
				bifurcation of nonlinear non-{H}ermitian skin effect},}\ }\href {\doibase
		10.1103/vxgf-59xt} {\bibfield  {journal} {\bibinfo  {journal} {Phys. Rev.
				Lett.}\ }\textbf {\bibinfo {volume} {135}},\ \bibinfo {pages} {126610}
		(\bibinfo {year} {2025})}\BibitemShut {NoStop}%
	\bibitem [{\citenamefont {Wu}\ \emph {et~al.}(2025)\citenamefont {Wu},
		\citenamefont {Hu}, \citenamefont {He}, \citenamefont {Deng}, \citenamefont
		{Huang}, \citenamefont {Ke}, \citenamefont {Deng}, \citenamefont {Lu},\ and\
		\citenamefont {Liu}}]{PhysRevLett.134.176601}%
	\BibitemOpen
	\bibfield  {author} {\bibinfo {author} {\bibfnamefont {J.}~\bibnamefont
			{Wu}}, \bibinfo {author} {\bibfnamefont {Y.}~\bibnamefont {Hu}}, \bibinfo
		{author} {\bibfnamefont {Z.}~\bibnamefont {He}}, \bibinfo {author}
		{\bibfnamefont {K.}~\bibnamefont {Deng}}, \bibinfo {author} {\bibfnamefont
			{X.}~\bibnamefont {Huang}}, \bibinfo {author} {\bibfnamefont
			{M.}~\bibnamefont {Ke}}, \bibinfo {author} {\bibfnamefont {W.}~\bibnamefont
			{Deng}}, \bibinfo {author} {\bibfnamefont {J.}~\bibnamefont {Lu}}, \ and\
		\bibinfo {author} {\bibfnamefont {Z.}~\bibnamefont {Liu}},\ }\bibfield
	{title} {\enquote {\bibinfo {title} {Hybrid-order skin effect from
				loss-induced nonreciprocity},}\ }\href {\doibase
		10.1103/PhysRevLett.134.176601} {\bibfield  {journal} {\bibinfo  {journal}
			{Phys. Rev. Lett.}\ }\textbf {\bibinfo {volume} {134}},\ \bibinfo {pages}
		{176601} (\bibinfo {year} {2025})}\BibitemShut {NoStop}%
	\bibitem [{\citenamefont {Jin}\ \emph {et~al.}(2025)\citenamefont {Jin},
		\citenamefont {Liu}, \citenamefont {Wang}, \citenamefont {Zhang},
		\citenamefont {Huang}, \citenamefont {Wei}, \citenamefont {Ju}, \citenamefont
		{Yang}, \citenamefont {Liu},\ and\ \citenamefont {Nori}}]{lpm2-vcb4}%
	\BibitemOpen
	\bibfield  {author} {\bibinfo {author} {\bibfnamefont {W.-W.}\ \bibnamefont
			{Jin}}, \bibinfo {author} {\bibfnamefont {J.}~\bibnamefont {Liu}}, \bibinfo
		{author} {\bibfnamefont {X.}~\bibnamefont {Wang}}, \bibinfo {author}
		{\bibfnamefont {Y.-R.}\ \bibnamefont {Zhang}}, \bibinfo {author}
		{\bibfnamefont {X.}~\bibnamefont {Huang}}, \bibinfo {author} {\bibfnamefont
			{X.}~\bibnamefont {Wei}}, \bibinfo {author} {\bibfnamefont {W.}~\bibnamefont
			{Ju}}, \bibinfo {author} {\bibfnamefont {Z.}~\bibnamefont {Yang}}, \bibinfo
		{author} {\bibfnamefont {T.}~\bibnamefont {Liu}}, \ and\ \bibinfo {author}
		{\bibfnamefont {F.}~\bibnamefont {Nori}},\ }\bibfield  {title} {\enquote
		{\bibinfo {title} {Anderson delocalization in strongly coupled disordered
				non-{H}ermitian chains},}\ }\href {\doibase 10.1103/lpm2-vcb4} {\bibfield
		{journal} {\bibinfo  {journal} {Phys. Rev. Lett.}\ }\textbf {\bibinfo
			{volume} {135}},\ \bibinfo {pages} {076602} (\bibinfo {year}
		{2025})}\BibitemShut {NoStop}%
	\bibitem [{\citenamefont {Jiang}\ \emph {et~al.}(2019)\citenamefont {Jiang},
		\citenamefont {Lang}, \citenamefont {Yang}, \citenamefont {Zhu},\ and\
		\citenamefont {Chen}}]{PhysRevB.100.054301}%
	\BibitemOpen
	\bibfield  {author} {\bibinfo {author} {\bibfnamefont {H.}~\bibnamefont
			{Jiang}}, \bibinfo {author} {\bibfnamefont {L.-J.}\ \bibnamefont {Lang}},
		\bibinfo {author} {\bibfnamefont {C.}~\bibnamefont {Yang}}, \bibinfo {author}
		{\bibfnamefont {S.-L.}\ \bibnamefont {Zhu}}, \ and\ \bibinfo {author}
		{\bibfnamefont {S.}~\bibnamefont {Chen}},\ }\bibfield  {title} {\enquote
		{\bibinfo {title} {Interplay of non-{H}ermitian skin effects and {A}nderson
				localization in nonreciprocal quasiperiodic lattices},}\ }\href {\doibase
		10.1103/PhysRevB.100.054301} {\bibfield  {journal} {\bibinfo  {journal}
			{Phys. Rev. B}\ }\textbf {\bibinfo {volume} {100}},\ \bibinfo {pages}
		{054301} (\bibinfo {year} {2019})}\BibitemShut {NoStop}%
	\bibitem [{\citenamefont {Liu}\ \emph {et~al.}(2020{\natexlab{a}})\citenamefont
		{Liu}, \citenamefont {Jiang}, \citenamefont {Cao},\ and\ \citenamefont
		{Chen}}]{PhysRevB.101.174205}%
	\BibitemOpen
	\bibfield  {author} {\bibinfo {author} {\bibfnamefont {Y.}~\bibnamefont
			{Liu}}, \bibinfo {author} {\bibfnamefont {X.-P.}\ \bibnamefont {Jiang}},
		\bibinfo {author} {\bibfnamefont {J.}~\bibnamefont {Cao}}, \ and\ \bibinfo
		{author} {\bibfnamefont {S.}~\bibnamefont {Chen}},\ }\bibfield  {title}
	{\enquote {\bibinfo {title} {Non-{H}ermitian mobility edges in
				one-dimensional quasicrystals with parity-time symmetry},}\ }\href {\doibase
		10.1103/PhysRevB.101.174205} {\bibfield  {journal} {\bibinfo  {journal}
			{Phys. Rev. B}\ }\textbf {\bibinfo {volume} {101}},\ \bibinfo {pages}
		{174205} (\bibinfo {year} {2020}{\natexlab{a}})}\BibitemShut {NoStop}%
	\bibitem [{\citenamefont {Longhi}(2019)}]{PhysRevLett.122.237601}%
	\BibitemOpen
	\bibfield  {author} {\bibinfo {author} {\bibfnamefont {S.}~\bibnamefont
			{Longhi}},\ }\bibfield  {title} {\enquote {\bibinfo {title} {Topological
				phase transition in non-{H}ermitian quasicrystals},}\ }\href {\doibase
		10.1103/PhysRevLett.122.237601} {\bibfield  {journal} {\bibinfo  {journal}
			{Phys. Rev. Lett.}\ }\textbf {\bibinfo {volume} {122}},\ \bibinfo {pages}
		{237601} (\bibinfo {year} {2019})}\BibitemShut {NoStop}%
	\bibitem [{\citenamefont {Tang}\ \emph {et~al.}(2021)\citenamefont {Tang},
		\citenamefont {Zhang}, \citenamefont {Zhang},\ and\ \citenamefont
		{Zhang}}]{PhysRevA.103.033325}%
	\BibitemOpen
	\bibfield  {author} {\bibinfo {author} {\bibfnamefont {L.-Z.}\ \bibnamefont
			{Tang}}, \bibinfo {author} {\bibfnamefont {G.-Q.}\ \bibnamefont {Zhang}},
		\bibinfo {author} {\bibfnamefont {L.-F.}\ \bibnamefont {Zhang}}, \ and\
		\bibinfo {author} {\bibfnamefont {D.-W.}\ \bibnamefont {Zhang}},\ }\bibfield
	{title} {\enquote {\bibinfo {title} {Localization and topological transitions
				in non-{H}ermitian quasiperiodic lattices},}\ }\href {\doibase
		10.1103/PhysRevA.103.033325} {\bibfield  {journal} {\bibinfo  {journal}
			{Phys. Rev. A}\ }\textbf {\bibinfo {volume} {103}},\ \bibinfo {pages}
		{033325} (\bibinfo {year} {2021})}\BibitemShut {NoStop}%
	\bibitem [{\citenamefont {Liu}\ \emph {et~al.}(2020{\natexlab{b}})\citenamefont
		{Liu}, \citenamefont {Guo}, \citenamefont {Pu},\ and\ \citenamefont
		{Longhi}}]{PhysRevB.102.024205}%
	\BibitemOpen
	\bibfield  {author} {\bibinfo {author} {\bibfnamefont {T.}~\bibnamefont
			{Liu}}, \bibinfo {author} {\bibfnamefont {H.}~\bibnamefont {Guo}}, \bibinfo
		{author} {\bibfnamefont {Y.}~\bibnamefont {Pu}}, \ and\ \bibinfo {author}
		{\bibfnamefont {S.}~\bibnamefont {Longhi}},\ }\bibfield  {title} {\enquote
		{\bibinfo {title} {Generalized {Aubry-Andr\'e} self-duality and mobility
				edges in non-{H}ermitian quasiperiodic lattices},}\ }\href {\doibase
		10.1103/PhysRevB.102.024205} {\bibfield  {journal} {\bibinfo  {journal}
			{Phys. Rev. B}\ }\textbf {\bibinfo {volume} {102}},\ \bibinfo {pages}
		{024205} (\bibinfo {year} {2020}{\natexlab{b}})}\BibitemShut {NoStop}%
	\bibitem [{\citenamefont {Weidemann}\ \emph {et~al.}(2022)\citenamefont
		{Weidemann}, \citenamefont {Kremer}, \citenamefont {Longhi},\ and\
		\citenamefont {Szameit}}]{Weidemann2022}%
	\BibitemOpen
	\bibfield  {author} {\bibinfo {author} {\bibfnamefont {S.}~\bibnamefont
			{Weidemann}}, \bibinfo {author} {\bibfnamefont {M.}~\bibnamefont {Kremer}},
		\bibinfo {author} {\bibfnamefont {S.}~\bibnamefont {Longhi}}, \ and\ \bibinfo
		{author} {\bibfnamefont {A.}~\bibnamefont {Szameit}},\ }\bibfield  {title}
	{\enquote {\bibinfo {title} {Topological triple phase transition in
				non-{H}ermitian {F}loquet quasicrystals},}\ }\href {\doibase
		10.1038/s41586-021-04253-0} {\bibfield  {journal} {\bibinfo  {journal}
			{Nature}\ }\textbf {\bibinfo {volume} {601}},\ \bibinfo {pages} {354}
		(\bibinfo {year} {2022})}\BibitemShut {NoStop}%
	\bibitem [{\citenamefont {Lin}\ \emph {et~al.}(2022)\citenamefont {Lin},
		\citenamefont {Li}, \citenamefont {Xiao}, \citenamefont {Wang}, \citenamefont
		{Yi},\ and\ \citenamefont {Xue}}]{Lin2022}%
	\BibitemOpen
	\bibfield  {author} {\bibinfo {author} {\bibfnamefont {Q.}~\bibnamefont
			{Lin}}, \bibinfo {author} {\bibfnamefont {T.}~\bibnamefont {Li}}, \bibinfo
		{author} {\bibfnamefont {L.}~\bibnamefont {Xiao}}, \bibinfo {author}
		{\bibfnamefont {K.}~\bibnamefont {Wang}}, \bibinfo {author} {\bibfnamefont
			{W.}~\bibnamefont {Yi}}, \ and\ \bibinfo {author} {\bibfnamefont
			{P.}~\bibnamefont {Xue}},\ }\bibfield  {title} {\enquote {\bibinfo {title}
			{Observation of non-{H}ermitian topological {A}nderson insulator in quantum
				dynamics},}\ }\href {\doibase 10.1038/s41467-022-30938-9} {\bibfield
		{journal} {\bibinfo  {journal} {Nat. Commun.}\ }\textbf {\bibinfo {volume}
			{13}},\ \bibinfo {pages} {3229} (\bibinfo {year} {2022})}\BibitemShut
	{NoStop}%
	\bibitem [{\citenamefont {Lin}\ \emph {et~al.}(2026)\citenamefont {Lin},
		\citenamefont {Cedzich}, \citenamefont {Zhou},\ and\ \citenamefont
		{Xue}}]{tz2n-lqxx}%
	\BibitemOpen
	\bibfield  {author} {\bibinfo {author} {\bibfnamefont {Q.}~\bibnamefont
			{Lin}}, \bibinfo {author} {\bibfnamefont {C.}~\bibnamefont {Cedzich}},
		\bibinfo {author} {\bibfnamefont {Q.}~\bibnamefont {Zhou}}, \ and\ \bibinfo
		{author} {\bibfnamefont {P.}~\bibnamefont {Xue}},\ }\bibfield  {title}
	{\enquote {\bibinfo {title} {Observation of metal-insulator and spectral
				phase transitions in {Aubry-Andr\'e-Harper} models},}\ }\href {\doibase
		10.1103/tz2n-lqxx} {\bibfield  {journal} {\bibinfo  {journal} {Phys. Rev.
				Lett.}\ }\textbf {\bibinfo {volume} {136}},\ \bibinfo {pages} {206602}
		(\bibinfo {year} {2026})}\BibitemShut {NoStop}%
	\bibitem [{\citenamefont {Liu}\ and\ \citenamefont
		{Chen}(2024)}]{PhysRevLett.133.193001}%
	\BibitemOpen
	\bibfield  {author} {\bibinfo {author} {\bibfnamefont {Y.}~\bibnamefont
			{Liu}}\ and\ \bibinfo {author} {\bibfnamefont {S.}~\bibnamefont {Chen}},\
	}\bibfield  {title} {\enquote {\bibinfo {title} {Fate of two-particle bound
				states in the continuum in non-{H}ermitian systems},}\ }\href {\doibase
		10.1103/PhysRevLett.133.193001} {\bibfield  {journal} {\bibinfo  {journal}
			{Phys. Rev. Lett.}\ }\textbf {\bibinfo {volume} {133}},\ \bibinfo {pages}
		{193001} (\bibinfo {year} {2024})}\BibitemShut {NoStop}%
	\bibitem [{\citenamefont {Huang}\ \emph {et~al.}(2024)\citenamefont {Huang},
		\citenamefont {Ke}, \citenamefont {Zhong}, \citenamefont {Kivshar},\ and\
		\citenamefont {Lee}}]{PhysRevLett.133.140202}%
	\BibitemOpen
	\bibfield  {author} {\bibinfo {author} {\bibfnamefont {B.}~\bibnamefont
			{Huang}}, \bibinfo {author} {\bibfnamefont {Y.}~\bibnamefont {Ke}}, \bibinfo
		{author} {\bibfnamefont {H.}~\bibnamefont {Zhong}}, \bibinfo {author}
		{\bibfnamefont {Y.~S.}\ \bibnamefont {Kivshar}}, \ and\ \bibinfo {author}
		{\bibfnamefont {C.}~\bibnamefont {Lee}},\ }\bibfield  {title} {\enquote
		{\bibinfo {title} {Interaction-induced multiparticle bound states in the
				continuum},}\ }\href {\doibase 10.1103/PhysRevLett.133.140202} {\bibfield
		{journal} {\bibinfo  {journal} {Phys. Rev. Lett.}\ }\textbf {\bibinfo
			{volume} {133}},\ \bibinfo {pages} {140202} (\bibinfo {year}
		{2024})}\BibitemShut {NoStop}%
	\bibitem [{\citenamefont {Bir}\ and\ \citenamefont {Pikus}(1974)}]{Bir1974}%
	\BibitemOpen
	\bibfield  {author} {\bibinfo {author} {\bibfnamefont {G.~L.}\ \bibnamefont
			{Bir}}\ and\ \bibinfo {author} {\bibfnamefont {G.}~\bibnamefont {Pikus}},\
	}\href@noop {} {\emph {\bibinfo {title} {Summetry and Strain-Induced Effects
				in Semiconductors}}}\ (\bibinfo  {publisher} {Keter, Jerusalem},\ \bibinfo
	{year} {1974})\BibitemShut {NoStop}%
	\bibitem [{\citenamefont {Cohen-Tannoudji}\ \emph {et~al.}(1998)\citenamefont
		{Cohen-Tannoudji}, \citenamefont {Dupont-Roc},\ and\ \citenamefont
		{Grynberg}}]{CCohenTannoudji1Atom}%
	\BibitemOpen
	\bibfield  {author} {\bibinfo {author} {\bibfnamefont {C.}~\bibnamefont
			{Cohen-Tannoudji}}, \bibinfo {author} {\bibfnamefont {J.}~\bibnamefont
			{Dupont-Roc}}, \ and\ \bibinfo {author} {\bibfnamefont {G.}~\bibnamefont
			{Grynberg}},\ }\href@noop {} {\emph {\bibinfo {title} {Atom-Photon
				Interactions}}}\ (\bibinfo  {publisher} {John Wiley and Sons},\ \bibinfo
	{year} {1998})\BibitemShut {NoStop}%
\end{thebibliography}

%

\end{document}